\documentclass[11pt,english]{article}
\usepackage[T1]{fontenc}
\usepackage[latin9]{inputenc}
\usepackage{verbatim}
\usepackage{float}
\usepackage{amsmath}
\usepackage{graphicx}
\usepackage{esint}
\usepackage{amsfonts}
\usepackage{hyperref}
\usepackage[numbers,sort&compress]{natbib}

\makeatletter

\providecommand{\tabularnewline}{\\}

\usepackage{latexsym}
\usepackage{epsfig}
\usepackage{enumerate}
\allowdisplaybreaks
\usepackage[numbers,sort&compress]{natbib}
\numberwithin{equation}{section}
\makeatother

\usepackage{babel}
\begin{document}
{}~ \hfill\vbox{\hbox{CTP-SCU/2013012}}\break \vskip 3.0cm
\centerline{\Large \bf New Cosmological Signatures from Double Field Theory}
\vspace*{10.0ex}

\begin{center}
 Houwen Wu and Haitang Yang
\end{center}

\vspace*{11.0ex}

\centerline{\large \it Center for theoretical physics,}
\centerline{\large \it College of Physical Science and Technology,}
\centerline{\large \it Sichuan University, Chengdu, 610064, China}

\vspace*{3.0ex}

\centerline{E-mail: 2013222020003@stu.scu.edu.cn, hyanga@scu.edu.cn}

\vspace*{10.0ex}

\centerline{\bf Abstract}
\bigskip

In cosmology, it has been a long-standing problem to establish a \emph{parameter insensitive} evolution from an anisotropic phase to an isotropic phase. On the other hand, it is of great importance to construct a theory having extra dimensions as its intrinsic ingredients.  We show that these two problems are closely related and can naturally be solved simultaneously in double field theory cosmology.
Our derivations are based on general arguments without any fine-tuning parameters.
In addition, We find that when we begin with FRW metric, the full spacetime metric of DFT totally agrees with \emph{Kaluza-Klein theory}. There is a visible and invisible dimension exchange
between the pre- and post-big bangs. Our results indicate that
double field theory has profound physical consequences and the continuous $O\left(D,D\right)$
is a very fundamental symmetry. This observation reinforces the viewpoint that symmetries dictate physics.

\smallskip

\vfill \eject \baselineskip=16pt \vspace*{10.0ex}

\tableofcontents


\section{Introduction}

Double field theory (DFT) \cite{Siegel:1993xq,Hull:2009mi,Tseytlin:1990nb, Duff:1989tf} is a renewed
formalism of closed string field theory. It doubles all spacetime
coordinates in order to make T-duality manifest on the level of component
fields. The usual coordinates $x^{i}$, conjugation of spacetime
momenta and its dual coordinates $\tilde{x}_{i}$, conjugation of
``winding'' numbers are treated on the same footing in this
theory. Then the fundamental coordinates are combined by the
$O\left(D,D\right)$ index
$X^{M}=\left(\tilde{x}_{i},x^{i}\right)$, where $M=1,2,\ldots,2D$
and $i=1,2,\ldots,D$. All component fields of closed string depend
on doubled coordinates $\phi\left(\tilde{x}_{i},x^{i}\right)$. Many progresses have been achieved based on DFT in recent years \cite{Zwiebach:2011rg} - \cite{Blumenhagen:2013zpa}.
In these works, Ref. \cite{Jeon:2010rw, Hohm:2010xe, Hohm:2011si, Vaisman, Vaisman2, Hohm:2012mf}
discussed the geometrical properties of DFT. Ref. \cite{Hohm:2011cp, Aldazabal:2011nj, Geissbuhler:2011mx, Grana:2012rr, Andriot:2012wx, Andriot:2012an, Geissbuhler:2013uka}
proposed scenarios of  relaxing the strong or weak constraints. Ref.
\cite{Kan:2011vg, Kan:2012nf, Wu:2013sha} are devoted to applications of  DFT on cosmology. Good reviews of DFT are referred to \cite{Zwiebach:2011rg,Aldazabal:2013sca,Berman:2013eva,Hohm:2013bwa}.

In string cosmology \cite{Gasperini:2007zz,Meissner:1991ge,Meissner:1991ge2,Tseytlin:1991xk,Gasperini:1991ak}, four types of solutions are constructed by the scale-factor duality and the
time-reversal transformation. These solutions  represent contracting or expanding universes, respectively.
Moreover, there are pre-big bang solutions which can not be constructed in the standard cosmology. It therefore states that the universe is not born from a singularity, but there exists
a pre-big bang phase. This is the main achievement of string cosmology in contrast to the standard cosmology. However, string cosmology suffers cosmic amnesia: The pre-big bang leaves no footprints in our post-big bang universe. Furthermore, the four solutions can be grouped arbitrarily in principle. One rescuing way is to introduce dilaton potentials to the theory. Though the cosmic amnesia and solution grouping problem can be alleviated, the universe has to be expanding or contracting all along the evolution, from the pre- to post-big bangs. For both experimental and theoretical reasons, it is widely accepted that the universe is anisotropic in the early stage. As a long-standing problem, string cosmology gives some explanations at the cost of free parameters and elaborate setups \cite{Batakis:1995kn,Chen:2001fh}.

In DFT cosmology, however, thanks to the intrinsic $O\left(D,D\right)$ symmetry, the pre- and post-big bangs, contracting and expanding universes are naturally unified in a single line element to cover the whole spacetime \cite{Wu:2013sha}.  With cross terms between two sets of coordinates $dxd\tilde{x}$, discussions on relations between two originally disconnected scenarios become possible. The purpose of this paper is to calculate solutions of DFT cosmology on the presence of all massless fields, and demonstrate some remarkable  cosmological implications and novel features. We  start with a constant Kalb-Ramond field

\begin{equation}
b_{ij}=\left(\begin{array}{cccc}
0 & c_{1} & c_{2} & c_{3}\\
-c_{1} & 0 & b_{1} & b_{2}\\
-c_{2} & -b_{1} & 0 & b_{3}\\
-c_{3} & -b_{2} & -b_{3} & 0
\end{array}\right),
\label{Const B}
\end{equation}
and an isotropic FRW like metric
\begin{equation}
g_{ij}=\left(\begin{array}{cccc}
-1 & 0 & 0 & 0\\
0 & a^{2}\left(\tilde{t},t\right) & 0 & 0\\
0 & 0 & a^{2}\left(\tilde{t},t\right) & 0\\
0 & 0 & 0 & a^{2}\left(\tilde{t},t\right)
\end{array}\right).
\label{Metric Ansatz}
\end{equation}
After substituting this metric  into the equations of motion (EOM) of the
DFT action and taking the synchronous gauge: $g_{tt}=-1$, $g_{ti}=b_{t\mu}=0$,
we get two metrics, the pre-big bang metric $dS_{pre}^{2}$
for $t<0$ and the post-big bang metric $dS_{post}^{2}$ for $t>0$.
Each of these metrics has two solutions, representing visibly isotropic or anisotropic evolutions respectively. The two metrics are disjointed by the big bang singularity,  which
implies that there is no interaction between the pre- and post-big bangs,
contracting and expanding universes. Therefore, there in principle are four choices of evolutions. However, it turns out that only one evolution reflects the evolution of our universe from simple physical consideration. In this physical evolution, the universe is explicitly isotropic only in the far future post-big bang region.

To explore the physics clearer, we further make use of the DFT action with an additional $O\left(D,D\right)$ invariant loop correction dilaton potential. We assume the dilaton potential
takes a form $e^{d}$, which represents higher loop corrections. This potential is defined to include a proper volume without breaking the generalized diffeomorphisms. Using
the $O\left(D,D\right)$ symmetry, we can obtain the solutions directly. Since the big bang singularity is smoothed out by the potential, the pre-big bang
metric $dS_{pre}^{2}$ and the post-big bang metric $dS_{post}^{2}$ are unified in
a single object  covering the full spacetime
backgrounds. It turns out the primary features do not rely on the potential.
Furthermore, after diagonalization, the metric can be rearranged as Kaluza-Klein ansatz. The dual coordinates become extra dimensions. If we compact these dimensions, the results totally agree with Kaluza-Klein compactification and the anti-symmetric Kalb-Ramond field decomposes into Kaluza-Klein vectors. Under the $O\left(D,D\right)$ symmetry, some dimensions will expand to visible and the others will contract to invisible.
These features are hidden in the traditional cosmology since there are no double coordinates.
Moreover, the existence of Kaluza-Klein extra dimensions in this theory is intrinsic but not artificial.

The reminder of this paper is outlined as follows. In section 2, we present the EOM of DFT. In section 3, we clarify the $O\left(D,D\right)$ symmetry and its applications in the traditional string cosmology and DFT.
In section 4, we put forward cosmological solutions of DFT and discuss their implications. Section 5 is our conclusion and discussions. In addition, appendix is devoted to detailed calculations to get the DFT solutions with vanishing potential.

\section{Equation of motion of double field theory}

The spacetime action of DFT  is built on the generalized $O\left(D,D\right)$ metric

\begin{equation}
\mathcal{H}_{MN}=\left(\begin{array}{cc}
g^{ij} & -g^{ik}b_{kj}\\
b_{ik}g^{kj} & g_{ij}-b_{ik}g^{kl}b_{lj}
\end{array}\right), \label{gen metric}
\end{equation}

\noindent which unifies the original spacetime metric $g_{ij}$ and the anti-symmetric
Kalb-Ramond field $b_{ij}$. The $O\left(D,D\right)$ indices $M$ and $N$ are raised and lowered by

\begin{equation}
\eta=\left(\begin{array}{cc}
0 & I\\
I & 0
\end{array}\right).
\end{equation}

\noindent The $O\left(D,D\right)$ invariant action is based on the generalized
metric

\begin{eqnarray}
S & = & \int dxd\tilde{x}e^{-2d}\left(\frac{1}{8}\mathcal{H}^{MN}\partial_{M}\mathcal{H}^{KL}\partial_{N}\mathcal{H}_{KL}-\frac{1}{2}\mathcal{H}^{MN}\partial_{N}\mathcal{H}^{KL}\partial_{L}\mathcal{H}_{MK}\right.\nonumber \\
 &  & \left.-\partial_{M}d\partial_{N}\mathcal{H}^{MN}+4\mathcal{H}^{MN}\partial_{M}d\partial_{N}d\right),
\end{eqnarray}

\noindent where the dilaton $d$ is an $O\left(D,D\right)$ scalar. The level matching condition in string theory leads to a constraint on the fields and gauge parameters  $\partial_{i}\tilde{\partial}^{i} \phi(t,\tilde t)=0$. The EOM of the dilaton is

\begin{eqnarray}
\frac{1}{8}\mathcal{H}^{MN}\partial_{M}\mathcal{H}^{KL}\partial_{N}\mathcal{H}_{KL}-
\frac{1}{2}\mathcal{H}^{MN}\partial_{M}\mathcal{H}^{KL}\partial_{K}\mathcal{H}_{NL}-
\partial_{M}\partial_{N}\mathcal{H}^{MN}\nonumber \\
-4\mathcal{H}^{MN}\partial_{M}d\partial_{N}d+4\partial_{M}
\mathcal{H}^{MN}\partial_{N}d+4\mathcal{H}^{MN}\partial_{M}\partial_{N}d  =  0,
\end{eqnarray}

\noindent with $\partial_{M}=\left(\tilde{\partial}^{i},\partial_{i}\right)$. The EOM of the generalized metric is

\begin{equation}
\mathcal{R}_{MN}\equiv\mathcal{K}_{MN}-S_{\quad M}^{P}\mathcal{K}_{PQ}S_{\quad N}^{Q} = 0,
\end{equation}

\noindent where

\begin{eqnarray}
\mathcal{K}_{MN} & = & \frac{1}{8}\partial_{M}\mathcal{H}^{KL}\partial_{N}\mathcal{H}_{KL}-\frac{1}{4}\left(\partial_{L}-2\partial_{L}d\right)\left(\mathcal{H}^{LK}\partial_{K}\mathcal{H}_{MN}\right)+2\partial_{M}\partial_{N}d\nonumber \\
 &  & -\frac{1}{2}\partial_{\left(N\right.}\mathcal{H}^{KL}\partial_{L}\mathcal{H}_{\left.M\right)K}+\frac{1}{2}\left(\partial_{L}-2\partial_{L}d\right)\left(\mathcal{H}^{KL}\partial_{\left(N\right.}\mathcal{H}_{\left.M\right)K}+\mathcal{H}_{\quad\left(M\right.}^{K}\partial_{K}\mathcal{H}_{\quad\left.N\right)}^{L}\right).
\end{eqnarray}

\noindent The most natural isotropic extension of FRW metric in DFT cosmology is
\begin{equation}
dS^{2}=g^{ij}d\tilde{x}_{i}d\tilde{x}_{j}-g^{ik}b_{kj}d\tilde{x}_{i}dx^{j}+b_{ik}g^{kj}dx^{i}
d\tilde{x}_{j}+\left(g_{ij}-b_{ik}g^{kl}b_{lj}\right)dx^{i}dx^{j},
\end{equation}

\noindent where the spacetime metric $g_{ij}$ is isotropic, as defined in eqn. (\ref{Metric Ansatz}). In this paper, we set $D=4$ and choose a constant Kalb-Ramond field as in eqn. (\ref{Const B})

\section{$O\left(D,D\right)$ transformations}
Before performing calculations, it is of importance to clarify some points about the symmetries. To our current understanding, the continuous $O\left(D,D\right)$ in DFT is a very fundamental symmetry.  Compactification of  $d=D-n$ dimensions breaks the continuous $O\left(D,D\right)$   into an $O\left(n,n\right)\times O\left(d,d;\mathbb{Z}\right)$ group. The $O\left(d,d;\mathbb{Z}\right)$  represents T-duality in the compactified background.  Also due to the $O\left(D,D\right)$, any background dependent on $\tilde x$ alone is also a solution of $x$ alone. This is the reason why the string  low energy effective action and supergravity  has the continuous $O\left(D,D\right)$ symmetry \cite{Sen:1991cn,Hassan:1991mq,Maharana:1992my,Sen:1994fa,Hull:1994ys}.  For this reason, the scale factor duality in string cosmology is not T-duality but a realization of  $O\left(D,D\right)$ symmetry in a specific set of backgrounds. As a bonus, this interpretation provides a simple way to construct solutions of DFT: two $O\left(D,D\right)$ paired solutions of the low energy effective action comprise a single DFT solution which  respects the constraint. Moreover, in the previous works \cite{Sen:1991cn,Meissner:1991ge,Sen:1991zi,Meissner:1991zj,Kar:1991kg,Sen:1992ua,Ali:1992mj},
the various solutions can be achieved by $O\left(D,D\right)$ transformations.
There are three types of generators of $O\left(D,D\right)$ transformations.
The first one belongs to $O\left(D\right)\times O\left(D\right)$ transformation,
which is a subgroup of $O\left(D,D\right)$. When we act the generators of $O\left(D\right)\times O\left(D\right)$
group on solutions, we get inequivalent new solutions. The $O\left(D\right)\times O\left(D\right)$
subgroup can be described by the following matrix

\begin{equation}
\Omega=\frac{1}{2}\left(\begin{array}{cc}
S+R & R-S\\
R-S & S+R
\end{array}\right), \label{O(d) trans}
\end{equation}

\noindent which satisfies $\Omega\eta\Omega^{T}=\eta$ and $S$, $R$
are $O\left(D\right)$ matrices. It is a maximal compact subgroup and
includes factorized duality \cite{Giveon:1994fu}. The new solutions
can be obtained by

\begin{equation}
M\longrightarrow\tilde{M}=\Omega^{T}M\Omega.
\end{equation}

\noindent The second subgroup generates  coordinate transformations and a shift of dilaton.
The matrix is

\noindent
\begin{equation}
\Omega=\left(\begin{array}{cc}
\left(A^{T}\right)^{-1} & 0\\
0 & A
\end{array}\right),\qquad A\in GL\left(D\right),
\end{equation}

\noindent where the constant matrix $A$ defines coordinate transformation
$x^{i}=A_{j}^{i}x^{j}$. When $A^{T}A=1$, it is included in the $O\left(D\right)\times O\left(D\right)$ transformation. The last
set of generators comes from gauge transformations, defined by the matrix

\noindent
\begin{equation}
\Omega=\left(\begin{array}{cc}
1 & 0\\
\Theta & 1
\end{array}\right),\label{eq:gauge trans}
\end{equation}

\noindent where elements of the anti-symmetric matrix $\Theta$ are constant parameters
of gauge transformations $b_{ij}\rightarrow b_{ij}+\Theta_{ij}$. If the spacetime is compact, the elements of these three generators should be integer numbers. Moreover, the last two sets of generators, which lie outside the
$O\left(D\right)\times O\left(D\right)$ group, do not generate new inequivalent solutions \cite{Sen:1991zi} in the traditional theory.

\subsection{$O\left(d,d\right)$ symmetry in string cosmology}

\noindent In the traditional string cosmology, coordinates are not
doubled and symmetry can be manifested in spatial parts of the metric. This is why
we only consider an $O\left(d,d\right)$ symmetry, but not an $O\left(D,D\right)$
symmetry in the traditional theory, where $d=D-1$. In the traditional spatial translation invariant string cosmology, there are two ways to manifest $O\left(d,d\right)$
symmetry. The first method is to find the
equations of motion of low energy effective action and the
EOM is invariant under $O\left(d,d\right)$ transformation. Another way
is to rewrite the action to make $O\left(d,d\right)$ symmetry
manifest. In this work, we adopt the second way, since it is
easy to figure out a relation between string cosmology and DFT.
The first one was reviewed in our previous work \cite{Wu:2013sha}.
We start by the low energy effective action

\noindent
\begin{equation}
S_{*}=\int d^{D}x\sqrt{-g}e^{-2\phi}\left[R+4\left(\partial_{\mu}\phi\right)^{2}-\frac{1}{12}H_{ijk}H^{ijk}\right],
\end{equation}

\noindent where $g_{\mu\nu}$ is the string metric, $\phi$ is the dilaton
and $H_{ijk}=3\partial_{\left[i\right.}b_{\left.jk\right]}$ is the field
strength of the anti-symmetric Kalb-Ramond $b_{ij}$ field. To consider the cosmological
backgrounds, we choose synchronous gauge $g_{tt}=-1$, $g_{ti}=b_{t\mu}=0$,
and define the $O\left(d,d\right)$ dilaton or shifted dilaton $d$ \footnote{Do not confuse the dilaton $d$ and number of spatial dimensions $d$} as follows

\noindent
\begin{equation}
e^{-2d}=\sqrt{g}e^{-2\phi}.
\end{equation}

\noindent The low energy effective action can be rewritten as

\noindent
\begin{equation}
S=-\int dte^{-2d}\left[\frac{1}{8}\mathrm{Tr}\left(\dot{M}\dot{M}^{-1}\right)+4\dot{d}^{2}\right], \label{ST action}
\end{equation}

\noindent where

\noindent
\begin{equation}
M=\left(\begin{array}{cc}
G^{-1} & -G^{-1}B\\
BG^{-1} & G-BG^{-1}B
\end{array}\right),\label{M}
\end{equation}

\noindent $G$ and $B$ are spatial parts of $g_{ij}\left(t\right)$
and $b_{ij}\left(t\right)$. Therefore, $M$ is a $2d$ by $2d$ matrix. Note that this action is invariant under
the $O\left(d,d\right)$ transformations

\noindent
\begin{equation}
d\longrightarrow d,\qquad M\longrightarrow\tilde{M}=\Omega^{T}M\Omega, \label{O(d,d) trans}
\end{equation}

\noindent where  $\Omega$ is a constant matrix, satisfying

\noindent
\begin{equation}
\Omega^{T}\eta\Omega=\eta,
\end{equation}

\noindent and $\eta$ is an invariant metric of the $O\left(d,d\right)$
group

\noindent
\begin{equation}
\eta=\left(\begin{array}{cc}
0 & I\\
I & 0
\end{array}\right).
\end{equation}

\noindent If we choose $G_{ij}=\delta_{ij}a^{2}\left(t\right)$ and $B=0$, the matrix (\ref{M}) becomes

\noindent
\begin{equation}
M=\left(\begin{array}{cc}
G^{-1} & 0\\
0 & G
\end{array}\right).
\end{equation}

\noindent After applying the $O(d)\times O(d)$ transformation (\ref{O(d) trans}), we have a new inequivalent solution

\noindent
\begin{equation}
\tilde{M}=\left(\begin{array}{cc}
G & 0\\
0 & G^{-1}
\end{array}\right),
\end{equation}

\noindent which implies that the action is invariant under $a\left(t\right)\rightarrow a^{-1}\left(t\right)$.
It is the so-called scale-factor duality in the traditional string cosmology.

\subsection{$O\left(D,D\right)$ symmetry in double field theory}

Now, let us consider the simplification of DFT action in terms of the matrix (\ref{M}). Recall the spacetime action of DFT

\noindent
\begin{equation}
S=\int d^{D}xd^{D}\tilde{x}\mathcal{L},
\end{equation}

\noindent where

\noindent
\begin{eqnarray}
\mathcal{L} & = & e^{-2d}\left(\frac{1}{8}\mathcal{H}^{MN}\partial_{M}\mathcal{H}^{KL}\partial_{N}\mathcal{H}_{KL}-\frac{1}{2}\mathcal{H}^{MN}\partial_{N}\mathcal{H}^{KL}\partial_{L}\mathcal{H}_{MK}\right.\nonumber \\
 &  & \left.-\partial_{M}d\partial_{N}\mathcal{H}^{MN}+4\mathcal{H}^{MN}\partial_{M}d\partial_{N}d\right).
\end{eqnarray}

\noindent This action is $O\left(D,D\right)$ invariant since each of
scalar terms is constructed by the contractions of $O\left(D,D\right)$
indices $M$ and $N$. It is a double formalism of the low energy
effective action. To see this, we expand each term as follows

\noindent
\begin{eqnarray}
\mathcal{L} & = & e^{-2d}\left[\frac{1}{8}\left(g_{ij}-b_{ik}g^{kl}b_{lj}\right)\tilde{\partial}^{i}\mathcal{H}^{KL}\tilde{\partial}^{j}\mathcal{H}_{KL}-\tilde{\partial}^{i}d\tilde{\partial}^{j}\left(g_{ij}-b_{ik}g^{kl}b_{lj}\right)+4\left(g_{ij}-b_{ik}g^{kl}b_{lj}\right)\tilde{\partial}^{i}d\tilde{\partial}^{j}d\right.\nonumber \\
 &  & +\frac{1}{8}b_{ik}g^{kj}\tilde{\partial}^{i}\mathcal{H}^{KL}\partial_{j}\mathcal{H}_{KL}-\tilde{\partial}^{i}d\partial_{j}\left(b_{ik}g^{kj}\right)+4\left(b_{ik}g^{kj}\right)\tilde{\partial}^{i}d\partial_{j}d\nonumber \\
 &  & +\frac{1}{8}\left(-g^{ik}b_{kj}\right)\partial_{i}\mathcal{H}^{KL}\tilde{\partial}^{j}\mathcal{H}_{KL}-\partial_{i}d\tilde{\partial}^{j}\left(-g^{ik}b_{kj}\right)+4\left(-g^{ik}b_{kj}\right)\partial_{i}d\tilde{\partial}^{j}d\nonumber \\
 &  & +\frac{1}{8}g^{ij}\partial_{i}\mathcal{H}^{KL}\partial_{j}\mathcal{H}_{KL}-\partial_{i}d\partial_{j}g^{ij}+4g^{ij}\partial_{i}d\partial_{j}d\nonumber \\
 &  & -\frac{1}{2}\left(g_{ij}-b_{im}g^{mn}b_{nj}\right)\tilde{\partial}^{j}\left(g_{kl}-b_{ka}g^{ab}b_{bl}\right)\tilde{\partial}^{l}g^{ik}-\frac{1}{2}\left(g_{ij}-b_{im}g^{mn}b_{nj}\right)\tilde{\partial}^{j}\left(b_{ka}g^{al}\right)\partial_{l}g^{ik}\nonumber \\
 &  & -\frac{1}{2}\left(g_{ij}-b_{im}g^{mn}b_{nj}\right)\tilde{\partial}^{j}\left(-g^{ka}b_{al}\right)\tilde{\partial}^{l}\left(-g^{ic}b_{ck}\right)-\frac{1}{2}\left(g_{ij}-b_{im}g^{mn}b_{nj}\right)\tilde{\partial}^{j}g^{kl}\partial_{l}\left(-g^{ic}b_{ck}\right)\nonumber \\
 &  & -\frac{1}{2}b_{im}g^{mj}\partial_{j}\left(g_{kl}-b_{ka}g^{ab}b_{bl}\right)\tilde{\partial}^{l}g^{ik}-\frac{1}{2}b_{im}g^{mj}\partial_{j}\left(b_{ka}g^{al}\right)\partial_{l}g^{ik}\nonumber \\
 &  & -\frac{1}{2}b_{im}g^{mj}\partial_{j}\left(-g^{ka}b_{al}\right)\tilde{\partial}^{l}\left(-g^{ic}b_{ck}\right)-\frac{1}{2}b_{im}g^{mj}\partial_{j}g^{kl}\partial_{l}\left(-g^{ic}b_{ck}\right)\nonumber \\
 &  & -\frac{1}{2}\left(-g^{im}b_{mj}\right)\tilde{\partial}^{j}\left(g_{kl}-b_{ka}g^{ab}b_{bl}\right)\tilde{\partial}^{l}\left(b_{ic}g^{ck}\right)-\frac{1}{2}\left(-g^{im}b_{mj}\right)\tilde{\partial}^{j}\left(b_{ka}g^{al}\right)\partial_{l}\left(b_{ic}g^{ck}\right)\nonumber \\
 &  & -\frac{1}{2}\left(-g^{im}b_{mj}\right)\tilde{\partial}^{j}\left(-g^{ka}b_{al}\right)\tilde{\partial}^{l}\left(g_{ik}-b_{ic}g^{cd}b_{dk}\right)-\frac{1}{2}\left(-g^{im}b_{mj}\right)\tilde{\partial}^{j}g^{kl}\partial_{l}\left(g_{ik}-b_{ic}g^{cd}b_{dk}\right)\nonumber \\
 &  & -\frac{1}{2}g^{ij}\partial_{j}\left(g_{kl}-b_{ka}g^{ab}b_{bl}\right)\tilde{\partial}^{l}\left(b_{ic}g^{ck}\right)-\frac{1}{2}g^{ij}\partial_{j}\left(b_{ka}g^{al}\right)\partial_{l}\left(b_{ic}g^{ck}\right)\nonumber \\
 &  & \left.-\frac{1}{2}g^{ij}\partial_{j}\left(-g^{ka}b_{al}\right)\tilde{\partial}^{l}\left(g_{ik}-b_{ic}g^{cd}b_{dk}\right)-\frac{1}{2}g^{ij}\partial_{j}g^{kl}\partial_{l}\left(g_{ik}-b_{ic}g^{cd}b_{dk}\right)\right).
\end{eqnarray}

\noindent The calculation rules can be found in our previous work \cite{Wu:2013sha}.
We assume all fields are double time dependent $g\left(\tilde{t},t\right)$
and also choose the synchronous gauge $g_{tt}=-1$, $g_{ti}=b_{t\mu}=0$.
The generalized Lagrangian can be simplified as

\begin{equation}
\mathcal{L}=e^{-2d}\left[-\frac{1}{8}\tilde{\partial}^{t}\mathcal{H}^{KL}\tilde{\partial}^{t}\mathcal{H}_{KL}-4\tilde{\partial}^{t}d\tilde{\partial}^{t}d-\frac{1}{8}\partial_{t}\mathcal{H}^{KL}\partial_{t}\mathcal{H}_{KL}-4\partial_{t}d\partial_{t}d\right],
\end{equation}

\noindent which can further reduce to

\begin{equation}
\mathcal{L}=e^{-2d}\left[-\frac{1}{8}\tilde{\partial}^{t}M^{KL}\tilde{\partial}^{t}M_{KL}
-4\dot{\tilde{d}}^{2}-\frac{1}{8}\partial_{t}M^{KL}\partial_{t}M_{KL}-4\dot{d}^{2}\right],\label{Lagran}
\end{equation}

\noindent with

\noindent
\begin{equation}
M=\left(\begin{array}{cc}
G^{-1} & -G^{-1}B\\
BG^{-1} & G-BG^{-1}B
\end{array}\right),
\end{equation}

\noindent where $\dot{d}\equiv\frac{\partial d}{\partial t}$ and
$\dot{\tilde{d}}\equiv\frac{\partial d}{\partial\tilde{t}}$. The metrics $G$ and $B$ are dependent
on two sets of coordinates, therefore $G$ and $B$ are spatial
parts of $g_{ij}\left(\tilde{t},t\right)$ and $b_{ij}\left(\tilde{t},t\right)$.
It is a difference to the traditional string cosmology. Furthermore, the first and third terms of Lagrangian (\ref{Lagran}) can be rewritten in a matrix representation

\noindent
\begin{equation}
S=-\int d^{D}xd^{D}\tilde{x}e^{-2d}\left[\frac{1}{8}\mathrm{Tr}\left(\dot{\tilde{M}}\dot{\tilde{M}}^{-1}\right)+4\dot{\tilde{d}}^{2}+\frac{1}{8}\mathrm{Tr}\left(\dot{M}\dot{M}^{-1}\right)+4\dot{d}^{2}\right],
\end{equation}

\noindent where $\dot{M}^{-1}\equiv\partial\left(M^{-1}\right)/\partial t$
and $\dot{\tilde{M}}^{-1}\equiv\partial\left(M^{-1}\right)/\partial \tilde{t}$.
To compare with (\ref{ST action}), it is obvious that this action  is also invariant
under $O\left(d,d\right)$ transformations: $M\longrightarrow\Omega^{T}M\Omega$
and $d\rightarrow d$. It is easy to see that
solutions of the traditional string cosmology are also solutions of
DFT.

\section{Cosmological solutions}
 In this section, we find virous solutions of DFT though $O(D,D)$ rotation. To respect the constraint, we adopt solutions dependent on only one set of coordinates, one time-like coordinate in particular in this paper. The term $\int d\tilde{t}$ can be integrated out. Therefore, the metric only includes one time-like direction $dt$.

\subsection{Solutions of DFT ($V\left(d\right)=0$) with constant Kalb-Ramond field}
\noindent There are two ways to approach solutions of DFT. The first method is to make use of the $O(D,D)$ symmetry. This method is easy though, one can only obtain solutions respecting constraints. The second way is to solve the EOM. It is possible to get other solutions by the second method, especially the constraint violating solutions. The detailed calculations of the second method are put in the Appendix and we give some constraint violating solutions there.

Making use of the $O(D,D)$ symmetry, the solutions of DFT for the isotropic FRW like metric are

\begin{eqnarray}
a_{\pm}\left(t\right)=\left(\frac{t}{t_{0}}\right)^{\pm1/\sqrt{D-1}}, & d\left(t\right)=-\frac{1}{2}\ln\frac{t}{t_{0}}, & t>0,\nonumber \\
a_{\pm}\left(-t\right)=\left(\frac{-t}{t_{0}}\right)^{\pm1/\sqrt{D-1}}, & d\left(-t\right)=-\frac{1}{2}\ln\frac{-t}{t_{0}}, & t<0,\label{eq:solution}
\end{eqnarray}

\noindent where $t_{0}$ is an initial time. In traditional string cosmology,
the physical interpretations of solutions (\ref{eq:solution}) are

\begin{center}
\begin{tabular}{|c|c|c|c|c|}
\hline
$a_{+}\left(t\right)$ & $\dot{a}\left(t\right)>0$, expansion & $\ddot{a}\left(t\right)<0$, decelerated & $\dot{H}<0$, decreasing curvature & post-big bang\tabularnewline
\hline
$a_{-}\left(t\right)$ & $\dot{\tilde{a}}\left(t\right)<0$, contraction & $\ddot{\tilde{a}}\left(t\right)>0$, decelerated & $\dot{\tilde{H}}>0$, decreasing curvature & post-big bang\tabularnewline
\hline
$a_{+}\left(-t\right)$ & $\dot{a}\left(-t\right)<0$, contraction & $\ddot{a}\left(-t\right)<0$, accelerated & $\dot{H}<0$, increasing curvature & pre-big bang\tabularnewline
\hline
$a_{-}\left(-t\right)$ & $\dot{\tilde{a}}\left(-t\right)>0$, expansion & $\ddot{\tilde{a}}\left(-t\right)>0$, accelerated & $\dot{\tilde{H}}>0$, increasing curvature & pre-big bang\tabularnewline
\hline
\end{tabular}

Table $1$. The properties of the solutions in tree level string
cosmology.
\par\end{center}

\noindent To get the solutions with constant Kalb-Ramond field, we could use $O(d,d)$ transformations (\ref{eq:gauge trans})

\begin{equation}
\Omega=\left(\begin{array}{cc}
1 & -B\\
0 & 1
\end{array}\right),
\end{equation}

\noindent If we consider $D=4$, $\Omega$ is 6 by 6 matrix and $B$ is an anti-symmetric constant matrix

\begin{equation}
B=\left(\begin{array}{ccc}
0 & b_{1} & b_{2}\\
-b_{1} & 0 & b_{3}\\
-b_{2} & -b_{3} & 0
\end{array}\right).
\end{equation}

\noindent Therefore, the new solution is
\begin{eqnarray}
\tilde{M} & = & \Omega^{T}M\Omega\nonumber \\
 & = & \left(\begin{array}{cc}
1 & 0\\
B & 1
\end{array}\right)\left(\begin{array}{cc}
G^{-1} & 0\\
0 & G
\end{array}\right)\left(\begin{array}{cc}
1 & -B\\
0 & 1
\end{array}\right)\nonumber \\
 & = & \left(\begin{array}{cc}
G^{-1} & -G^{-1}B\\
BG^{-1} & G-BG^{-1}B
\end{array}\right),
\end{eqnarray}

\noindent where $G_{ij}=\delta_{ij}a_{\pm}^{2}\left(t\right)$. Due to the singularity at $t=0$,  the line element splits into  two parts, one for the regime of $t<0$ and the other for $t>0$.

\noindent
\begin{eqnarray}
dS_{pre\pm}^{2} & = & -dt^{2}+a_\pm\left(-t\right)^{-2}\left(d\tilde{x}_{2}^{2}+d\tilde{x}_{3}^{2}+d\tilde{x}_{4}^{2}\right)+a_{\pm}\left(-t\right)^{2}\left(dx_{2}^{2}+dx_{3}^{2}+dx_{4}^{2}\right)\nonumber \\
 &  & -2a_{\pm}\left(-t\right)^{-2}b_{1}d\tilde{x}_{2}dx_{3}-2a_{\pm}\left(-t\right)^{-2}b_{2}d\tilde{x}_{2}dx_{4}-2a_{\pm}\left(-t\right)^{-2}b_{3}d\tilde{x}_{3}dx_{4}\nonumber \\
 &  & +2a_{\pm}\left(-t\right)^{-2}b_{1}d\tilde{x}_{3}dx_{2}+2a_{\pm}\left(-t\right)^{-2}b_{2}d\tilde{x}_{4}dx_{2}+2a_{\pm}\left(-t\right)^{-2}b_{3}d\tilde{x}_{4}dx_{3}\nonumber \\
 &  & +a_{\pm}\left(-t\right)^{-2}\left(b_{1}^{2}+b_{2}^{2}\right)dx_{2}^{2}+a_{\pm}\left(-t\right)^{-2}\left(b_{1}^{2}+b_{3}^{2}\right)dx_{3}^{2}+a_{\pm}\left(-t\right)^{-2}\left(b_{2}^{2}+b_{3}^{2}\right)dx_{4}^{2}\nonumber \\
 &  & +2a_{\pm}\left(-t\right)^{-2}b_{1}b_{2}dx_{4}dx_{3}-2a_{\pm}\left(-t\right)^{-2}b_{1}b_{3}dx_{4}dx_{2}+2a_{\pm}\left(-t\right)^{-2}b_{2}b_{3}dx_{3}dx_{2},\qquad t<0,\label{eq:pre metric}\\
\nonumber \\
dS_{post\pm}^{2} & = & -dt^{2}+a_\pm\left(t\right)^{-2}\left(d\tilde{x}_{2}^{2}+d\tilde{x}_{3}^{2}+d\tilde{x}_{4}^{2}\right)+a_\pm\left(t\right)^{2}\left(dx_{2}^{2}+dx_{3}^{2}+dx_{4}^{2}\right)\nonumber \\
 &  & -2a_\pm\left(t\right)^{-2}b_{1}d\tilde{x}_{2}dx_{3}-2a_\pm\left(t\right)^{-2}b_{2}d\tilde{x}_{2}dx_{4}-2a_\pm\left(t\right)^{-2}b_{3}d\tilde{x}_{3}dx_{4}\nonumber \\
 &  & +2a_\pm\left(t\right)^{-2}b_{1}d\tilde{x}_{3}dx_{2}+2a_\pm\left(t\right)^{-2}b_{2}d\tilde{x}_{4}dx_{2}+2a_\pm\left(t\right)^{-2}b_{3}d\tilde{x}_{4}dx_{3}\nonumber \\
 &  & +a_\pm\left(t\right)^{-2}\left(b_{1}^{2}+b_{2}^{2}\right)dx_{2}^{2}+a_\pm\left(t\right)^{-2}\left(b_{1}^{2}+b_{3}^{2}\right)dx_{3}^{2}+a_\pm\left(t\right)^{-2}\left(b_{2}^{2}+b_{3}^{2}\right)dx_{4}^{2}\nonumber \\
 &  & +2a_\pm\left(t\right)^{-2}b_{1}b_{2}dx_{4}dx_{3}-2a_\pm\left(t\right)^{-2}b_{1}b_{3}dx_{4}dx_{2}+2a_\pm\left(t\right)^{-2}b_{2}b_{3}dx_{3}dx_{2},\qquad t>0.\label{eq:post metric}
\end{eqnarray}

\noindent The strong constraint violated solution can be found in Appendix.

\subsection{Solutions of DFT ($V\left(d\right)\neq0$) with constant Kalb-Ramond field}
To understand the physics better, we start with the action
\begin{equation}
S=\int dx^{D}d\tilde{x}^{D}e^{-2d}\left[\mathcal{R}+V\left(d\right)\right],
\label{Dilaton Potential S}
\end{equation}

\noindent where $V\left(d\right)$ is an $O\left(D,D\right)$ invariant potential

\begin{equation}
V\left(d\right)=V_{0}e^{8d}.
\end{equation}

\noindent This non-local potential represents the backreaction of higher loop corrections and $V_{0}>0$ \cite{Gasperini:1996np}. It is worth  noting that this dilaton has been redefined to include a proper volume which will reduce to $d(t,\tilde t)$ in cosmological background. It leads the dilaton potential above to be a scalar under generalized diffeomorphisms. Utilizing the $O\left(D,D\right)$ symmetry, one can figure out a DFT cosmological solution without Kalb-Ramond field is
\begin{equation}
a\left(t\right)=a_{0}\left[\frac{t}{t_{0}}+\left(1+\frac{t^{2}}{t_{0}^{2}}\right)^{\frac{1}{2}}\right]^{\frac{1}{\sqrt{3}}},\qquad d\left(t\right)=-\frac{1}{4}\ln\left[\sqrt{V_{0}}t_{0}\left(1+\frac{t^{2}}{t_{0}^{2}}\right)\right],
\label{string cosmology solution}
\end{equation}

\noindent where $t_{0}$ and $V_{0}$ are integral constants. It is of interest to note that there is one single solution of the scale factor.
A single metric unifies the pre- and post- big bangs, expanding and contracting universes without any singularity. The metric is

\begin{equation}
ds^{2}=-dt^{2}+a^{-2}\left(d\tilde{x}_{2}^{2}+d\tilde{x}_{3}^{2}+d\tilde{x}_{4}^{2}\right)+a^{2}\left(dx_{2}^{2}+dx_{3}^{2}+dx_{4}^{2}\right).\label{eq:sol b=0}
\end{equation}

\noindent To obtain the solutions with constant Kalb-Ramond field, we also choose the transformation matrix as

\begin{equation}
\Omega=\left(\begin{array}{cc}
1 & -B\\
0 & 1
\end{array}\right).
\end{equation}

\noindent After applying $\tilde{M}=\Omega^{T}M\Omega$, the metric becomes

\begin{eqnarray}
dS^{2} & = & -dt^{2}+a\left(t\right)^{-2}\left(d\tilde{x}_{2}^{2}+d\tilde{x}_{3}^{2}+d\tilde{x}_{4}^{2}\right) +a\left(t\right)^{2}\left(dx_{2}^{2}+dx_{3}^{2}+dx_{4}^{2}\right)\nonumber \\
 &  & -2a\left(t\right)^{-2}b_1d\tilde{x}_{2}dx_{3}-2a\left(t\right)^{-2}b_2 d\tilde{x}_{2}dx_{4}-2a\left(t\right)^{-2}b_3 d\tilde{x}_{3}dx_{4}\nonumber \\
 &  &+2a\left(t\right)^{-2}b_1d\tilde{x}_{3}dx_{2} +2a\left(t\right)^{-2}b_2d\tilde{x}_{4}dx_{2}+2a\left(t\right)^{-2}b_3d\tilde{x}_{4}dx_{3}\nonumber \\
 & & + a(t)^{-2} (b_1^2+b_2^2)dx_2^2 + a(t)^{-2} (b_1^2+b_3^2)dx_3^2 +a(t)^{-2} (b_2^2+b_3^2)dx_4^2\nonumber \\
 &  & +2a\left(t\right)^{-2}b_1 b_{2}dx_{4}dx_{3}-2a\left(t\right)^{-2}b_1 b_3dx_{4}dx_{2}+2a\left(t\right)^{-2}b_{2}b_3dx_{3}dx_{2}.
\end{eqnarray}

\noindent This metric can be rearranged as

\begin{eqnarray}
ds^{2} & = & -dt^{2}+a^{-2}\left(d\tilde{x}_{2}-b_{1}dx_{3}-b_{2}dx_{4}\right)^{2}+a^{-2}\left(d\tilde{x}_{3}+b_{1}dx_{2}-b_{3}dx_{4}\right)^{2}\nonumber \\
 &  & +a^{-2}\left(d\tilde{x}_{4}+b_{2}dx_{2}+b_{3}dx_{3}\right)^{2}+a^{2}\left(dx_{2}^{2}+dx_{3}^{2}+dx_{4}^{2}\right),
\end{eqnarray}

\noindent One can see that  coordinates $\tilde{x}$ have the same form as Kaluza-Klein extra dimensions.
To make it clear, we use $\alpha$, $\beta$ $\left(=5,6,7\right)$ to denote the indices
of extra dimensions and define $x^{5}\equiv\tilde{x}^{2}$, $x^{6}\equiv\tilde{x}^{3}$,
$x^{7}\equiv\tilde{x}^{4}$ . Therefore, the metric can be recasted into the form

\begin{equation}
ds^{2}=g_{\mu\nu}dx^{\mu}dx^{\nu}+g_{\alpha\beta}\left(dx^{\alpha}+A_{\mu}^{\alpha}dx^{\mu}\right)\left(dx^{\beta}+A_{\mu}^{\beta}dx^{\mu}\right),\label{KK}
\end{equation}

\noindent with

\begin{eqnarray}
A_{\mu}^{5} & = & \left(0,0,-b_{1},-b_{2}\right),\nonumber \\
A_{\mu}^{6} & = & \left(0,b_{1},0,-b_{3}\right),\nonumber \\
A_{\mu}^{7} & = & \left(0,b_{2},b_{3},0\right),\nonumber \\
g_{\mu\nu} & = & \mathrm{diag}\left(-1,a^{2},a^{2},a^{2}\right),\nonumber \\
g_{\alpha\beta} & = & \mathrm{diag}\left(a^{-2},a^{-2},a^{-2}\right),
\end{eqnarray}

\noindent where $dx^{\alpha}$ are extra dimensions, all fields do not depend on $x^{\alpha}$, $g_{\alpha\beta}$ are radii of extra dimensions and $A_{\mu}^{\alpha}$ are Kaluza-Klein vectors. If we impose the periodic condition $x^{\alpha}\sim x^{\alpha}+2 \pi R$ and choose integer number $b_{i}$, this metric is in total agreement with Kaluza-Klein compactification.

\noindent
\begin{figure}[H]
\begin{centering}
\includegraphics[scale=0.7]{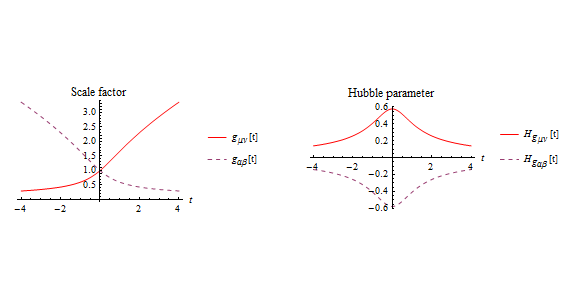}
\par\end{centering}
\caption{Scale factors and Hubble parameters}
\label{S and H}
\end{figure}

\noindent The scale factors and relevant Hubble parameters are plotted in FIG. (\ref{S and H}). It is easy to see that the visible dimensions contract and the extra dimensions expand in the pre-big bang. As time evolves, The visible dimensions will contract to invisible dimensions and extra dimensions become visible in the post-big bang. In other words, the universe  evolves smoothly from an anisotropic phase  to an isotropic phase, and again to anisotropy.
This observation proposes important implications for future experiments. Once extra
dimensions are detected, we can expect information of the pre-big bang could be extracted, which is of
help for the cosmic amnesia puzzle. A more speculating application is that we can compare the size of
the extra dimension with the scale of the universe. Our calculation indicates that one is the reciprocal
of the other. It is reasonable that the real story is more complicated since our discussion in this work
does not contain massive contributions and quantum effects. Nevertheless, we do have another way to
justify string theory and the existence of the pre-big bang.
Moreover, the whole process is even more parameter insensitive! Bear in mind we do
not need to assume any anisotropy at the very beginning for the metric entries.  Comparing with the metric of $b=0$ (\ref{eq:sol b=0}), we find a duality between two distinct spacetime backgrounds

\begin{eqnarray}
b=0 &  & b\neq0\nonumber \\
\tilde{x}_{2} & \longleftrightarrow & \tilde{x}_{2}-b_{1}x_{3}-b_{2}x_{4}\nonumber \\
\tilde{x}_{3} & \longleftrightarrow & \tilde{x}_{3}+b_{1}x_{2}-b_{3}x_{4}\nonumber \\
\tilde{x}_{4} & \longleftrightarrow & \tilde{x}_{4}+b_{2}x_{2}+b_{3}x_{3}.
\end{eqnarray}

\noindent Now, let us clarify the differences between solutions of the traditional
string cosmology and DFT. In the traditional string cosmology, the
rotation matrix $\Omega$ is treated as gauge transformations  (\ref{eq:gauge trans}).
It does not generate new inequivalent solutions. For example, let us begin with
$4-$dimensional isotropic FRW metric in string frame, where

\noindent
\begin{equation}
ds^{2}=-dt^{2}+a\left(t\right)^{2}\left(dx_{2}^{2}+dx_{3}^{2}+dx_{4}^{2}\right).\label{FRW}
\end{equation}

\noindent In the $O\left(d,d\right)$ language, the matrix $G$ is

\noindent
\begin{equation}
G=\left(\begin{array}{ccc}
a{}^{2} & 0 & 0\\
0 & a{}^{2} & 0\\
0 & 0 & a{}^{2}
\end{array}\right),\qquad B=0,
\end{equation}

\noindent After rotation
$M\longrightarrow\tilde{M}=\Omega^{T}M\Omega$, the new solution
is

\noindent
\begin{equation}
\tilde{G}=\left(\begin{array}{ccc}
a{}^{2} & 0 & 0\\
0 & a{}^{2} & 0\\
0 & 0 & a{}^{2}
\end{array}\right),\qquad\tilde{B}=\left(\begin{array}{ccc}
0 & b_{1} & b_{2}\\
-b_{1} & 0 & b_{3}\\
-b_{2} & -b_{3} & 0
\end{array}\right).
\end{equation}

\noindent However, this new solution gives us the same  spacetime background,
since $\tilde{G}_{ij}=G_{ij}=\delta_{ij}a^{2}\left(t\right)$. It
means that the rotation gives no  new inequivalent solution but is only a gauge
transformation of $B$, shifting the Kalb-Ramond field from $0$ to
$\tilde{B}_{ij}$. Since $\tilde{B}$ is a constant matrix, it is
a pure gauge.

In DFT, since we double all coordinates, when we consider the spacetime
background, we can not  consider the matrix $G$ (representing one set of coordinates) only, but a full matrix $M$. Now, we
also start with same metric (\ref{FRW}) as
\begin{equation}
ds^{2}=-dt^{2}+a\left(t\right)^{2}\left(dx_{2}^{2}+dx_{3}^{2}+dx_{4}^{2}\right).
\end{equation}

\noindent The $O\left(d,d\right)$ symmetry gives us the full spacetime
metric

\begin{equation}
ds^{2}=-dt^{2}+a^{-2}\left(d\tilde{x}_{2}^{2}+d\tilde{x}_{3}^{2}+d\tilde{x}_{4}^{2}\right)+a^{2}\left(dx_{2}^{2}+dx_{3}^{2}+dx_{4}^{2}\right).
\end{equation}

\noindent The relevant matrix $M$ is

\noindent
\begin{equation}
M=\left(\begin{array}{cccccc}
a{}^{-2} & 0 & 0 & 0 & 0 & 0\\
0 & a{}^{-2} & 0 & 0 & 0 & 0\\
0 & 0 & a{}^{-2} & 0 & 0 & 0\\
0 & 0 & 0 & a{}^{2} & 0 & 0\\
0 & 0 & 0 & 0 & a{}^{2} & 0\\
0 & 0 & 0 & 0 & 0 & a{}^{2}
\end{array}\right).\label{new M0}
\end{equation}

\noindent Utilizing $O\left(d,d\right)$ transformation by matrix
$\Omega$, a new solution is generated

\begin{equation}
\tilde{M}=\left(\begin{array}{cccccc}
a{}^{-2} & 0 & 0 & 0 & -b_{1}a^{-2} & -b_{2}a^{-2}\\
0 & a{}^{-2} & 0 & b_{1}a^{-2} & 0 & -b_{3}a^{-2}\\
0 & 0 & a{}^{-2} & b_{2}a^{-2} & b_{3}a^{-2} & 0\\
0 & b_{1}a^{-2} & b_{2}a^{-2} & \left(b_{1}^{2}+b_{2}^{2}\right)a^{-2}+a{}^{2} & b_{2}b_{3}a^{-2} & -b_{1}b_{3}a^{-2}\\
-b_{1}a^{-2} & 0 & b_{3}a^{-2} & b_{2}b_{3}a^{-2} & \left(b_{1}^{2}+b_{3}^{2}\right)a^{-2}+a{}^{2} & b_{1}b_{2}a^{-2}\\
-b_{2}a^{-2} & -b_{3}a^{-2} & 0 & -b_{1}b_{3}a^{-2} & b_{1}b_{2}a^{-2} & \left(b_{2}^{2}+b_{3}^{2}\right)a^{-2}+a{}^{2}
\end{array}\right).\label{new M}
\end{equation}

\noindent Although, the part of $G$ in $M$ (\ref{new M0}) and the part of $\tilde{G}$ in $\tilde M$ (\ref{new M}) are the same with each
other (same solutions), the full matrix $M$ is different with  $\tilde{M}$.
Therefore, we can argue that this solution (\ref{new M}) has a new inequivalent
physical interpretation in cosmological background. This is a difference between
the traditional string cosmology and DFT when we begin with the same metrics
and utilize the same rotation.

\section{Conclusion}
In summary, in this paper, we first clarified that  T-duality in compactified background and $O(D,D)$ symmetry in low energy effective action are descents of the continuous $O(D,D)$ symmetry of DFT. Next, we present the differences and relations between the traditional string cosmology and DFT in terms of $O(d,d)$ matrix (\ref{M}).
We discussed the cosmological interpretation of DFT. To  understand the physics better, we introduced an $O(D,D)$ invariant dilaton potential, derived from the backreaction of higher loop corrections, into the DFT action to calculate isotropic FRW like solutions. Our results exhibit some remarkable novel properties:

\begin{itemize}
\item Extra dimensions naturally arise in DFT cosmology without any elaborate pre-assumptions, it takes a form of Kaluza-Klein compactification.

\item The original equivalent solutions in the traditional string cosmology have inequivalent physical interpretation in DFT, since there are mixed terms of double coordinates.

\item Though starting from an isotropic FRW like metric, surprisingly, the universe evolves smoothly from a clearly anisotropic phase all the way to an visible isotropic phase at the big bang. And again to an anisotropic phase as time goes by.

\item The whole evolution is parameter insensitive. No initial condition is needed. No initial anisotropy assumption is required.

\item In the region $t\to -\infty$, there are extra dimensions hidden by the weak string coupling.  While in the region $t\sim 0$, all dimensions are revealed and visible. Visible dimensions in the pre-big bang are hidden as $t\to \infty$.

\item The visible/hidden dimension exchange sheds some light on the exploration of the pre-big bang and cosmic amnesia puzzle. It also proposes a possible justification of string theory. To fulfill these purposes, further researches are needed.

\end{itemize}
In brief, not only does DFT provide an aesthetic formalism, it also has profound physical contents. It is immediately of great interest to address the influence on particle physics by spacetime backgrounds (\ref{KK}). 

\vspace{5mm}

\noindent {\bf Acknowledgements. }
We would like to acknowledge illuminating discussions with T. Li, J. Lu, Z. Sun and P. Wang. This work is supported in part by the NSFC (Grant No. 11175039 and 11375121
) and SiChuan Province Science Foundation for Youths (Grant No. 2012JQ0039). H. Y. is grateful to the hospitality of the Institute of Theoretical Physics, Chinese Academy of Sciences where part of this work is done.

\newpage

\appendix

\section{Appendix}

\subsection{Some notations and calculation rules}

The generalized metric and its inverse are given by

\begin{equation}
\mathcal{H}_{MN}=\left(\begin{array}{cc}
g^{ij} & -g^{ik}b_{kj}\\
b_{ik}g^{kj} & g_{ij}-b_{ik}g^{kl}b_{lj}
\end{array}\right),\quad\mathcal{H}^{MN}=\left(\begin{array}{cc}
g_{ij}-b_{ik}g^{kl}b_{lj} & b_{ik}g^{kj}\\
-g^{ik}b_{kj} & g^{ij}
\end{array}\right).
\end{equation}

\noindent The inner product of coordinates and their duals are defined as

\noindent
\begin{eqnarray}
\mathcal{H}_{\mathbf{1}\mathbf{1}}\left(\frac{\partial}{\partial\tilde{x}_{i}},\frac{\partial}{\partial\tilde{x}_{j}}\right)=g^{ij}, &  & \mathcal{H}_{\mathbf{1}\mathbf{2}}\left(\frac{\partial}{\partial\tilde{x}_{i}},\frac{\partial}{\partial x^{j}}\right)=-g^{ik}b_{kj},\nonumber \\
\mathcal{H}_{\mathbf{2}\mathbf{1}}\left(\frac{\partial}{\partial x^{i}},\frac{\partial}{\partial\tilde{x}_{j}}\right)=b_{ik}g^{kj}, &  & \mathcal{H}_{\mathbf{\mathbf{2}}\mathbf{2}}\left(\frac{\partial}{\partial x^{i}},\frac{\partial}{\partial x^{j}}\right)=g_{ij}-b_{ik}g^{kl}b_{lj}.
\end{eqnarray}

\noindent Making use of following new notations to represent the block
metrix of generalized metric

\noindent
\begin{eqnarray}
\mathcal{H}_{\mathbf{1}\left(i\right)\mathbf{1}\left(j\right)}=\mathcal{G}^{ij}, &  & \mathcal{H}^{\mathbf{1}\left(i\right)\mathbf{1}\left(j\right)}=\mathcal{G}_{ij},\nonumber \\
\mathcal{H}_{\mathbf{1}\left(i\right)\mathbf{2}\left(j\right)}=\mathcal{B}_{\; j}^{i}, &  & \mathcal{H}^{\mathbf{1}\left(i\right)\mathbf{2}\left(j\right)}=\mathcal{B}_{i}^{\; j},\nonumber \\
\mathcal{H}_{\mathbf{2}\left(i\right)\mathbf{1}\left(j\right)}=\mathcal{C}_{i}^{\; j}, &  & \mathcal{H}^{\mathbf{2}\left(i\right)\mathbf{1}\left(j\right)}=\mathcal{C}_{\; j}^{i},\nonumber \\
\mathcal{H}_{\mathbf{\mathbf{2}}\left(i\right)\mathbf{2}\left(j\right)}=\mathcal{D}_{ij}, &  & \mathcal{H}^{\mathbf{\mathbf{2}}\left(i\right)\mathbf{2}\left(j\right)}=\mathcal{D}^{ij},
\end{eqnarray}

\noindent and

\noindent
\begin{eqnarray}
\mathcal{H}_{\quad\mathbf{1}}^{M} & = & \mathcal{H}^{M\mathbf{2}}\eta_{\mathbf{2}\mathbf{1}}=\mathcal{H}^{M\mathbf{2}},\nonumber \\
\mathcal{H}_{\quad\mathbf{2}}^{M} & = & \mathcal{H}^{M\mathbf{1}}\eta_{\mathbf{1}\mathbf{2}}=\mathcal{H}^{M\mathbf{1}},
\end{eqnarray}

\noindent we find

\noindent
\begin{eqnarray}
\mathcal{H}_{\quad\mathbf{1}\left(j\right)}^{\mathbf{1}\left(i\right)}=\mathcal{H}^{\mathbf{1}\left(i\right)\mathbf{2}\left(j\right)}=\mathcal{B}_{i}^{\; j} &  & \mathcal{H}_{\quad\mathbf{2}\left(j\right)}^{\mathbf{1}\left(i\right)}=\mathcal{H}^{\mathbf{1}\left(i\right)\mathbf{1}\left(j\right)}=\mathcal{G}_{ij},\nonumber \\
\mathcal{H}_{\quad\mathbf{1}\left(j\right)}^{\mathbf{2}\left(i\right)}=\mathcal{H}^{\mathbf{2}\left(i\right)\mathbf{2}\left(j\right)}=\mathcal{D}^{ij} &  & \mathcal{H}_{\quad\mathbf{2}\left(j\right)}^{\mathbf{2}\left(i\right)}=\mathcal{H}^{\mathbf{2}\left(i\right)\mathbf{1}\left(j\right)}=\mathcal{C}_{\; j}^{i}.
\end{eqnarray}

\noindent Then  the background metrics are
forms

\noindent
\begin{equation}
\tilde{g}^{ij}=\left(\begin{array}{cccc}
-1\\
 & \tilde{a}{}^{2}\\
 &  & \tilde{a}{}^{2}\\
 &  &  & \tilde{a}{}^{2}
\end{array}\right),\qquad g_{ij}=\left(\begin{array}{cccc}
-1\\
 & a^{2}\\
 &  & a^{2}\\
 &  &  & a^{2}
\end{array}\right),
\end{equation}

\noindent where

\noindent
\begin{equation}
\tilde{a}=a^{-1}
\end{equation}

\noindent In calculations, we only have two combinations of derivatives
and metrics. The detailed calculation rules can be read in \cite{Wu:2013sha}

\noindent
\begin{equation}
g^{\bullet\bullet}\partial g_{\bullet\bullet},\qquad\tilde{g}^{\bullet\bullet}\tilde{\partial}\tilde{g}_{\bullet\bullet},
\end{equation}

\noindent where we choose the simplest $b$ field

\noindent
\begin{equation}
b_{ij}=\left(\begin{array}{cccc}
0 & 0 & 0 & 0\\
0 & 0 & 0 & 0\\
0 & 0 & 0 & b\\
0 & 0 & -b & 0
\end{array}\right),
\end{equation}

\noindent with $b$ is a non-zero constant.

\subsection{Equation of motion of the dilaton}

\begin{eqnarray}
 &  & \frac{1}{8}\mathcal{H}^{MN}\partial_{M}\mathcal{H}^{KL}\partial_{N}\mathcal{H}_{KL}-\frac{1}{2}\mathcal{H}^{MN}\partial_{M}\mathcal{H}^{KL}\partial_{K}\mathcal{H}_{NL}\nonumber \\
 &  & -\partial_{M}\partial_{N}\mathcal{H}^{MN}-4\mathcal{H}^{MN}\partial_{M}d\partial_{N}d+4\partial_{M}\mathcal{H}^{MN}\partial_{N}d+4\mathcal{H}^{MN}\partial_{M}\partial_{N}d\nonumber \\
 & = & \frac{1}{8}\mathcal{H}^{\mathbf{1}\left(m\right)\mathbf{1}\left(n\right)}\partial_{\mathbf{1}\left(m\right)}\mathcal{H}^{\mathbf{1}\left(i\right)\mathbf{1}\left(j\right)}\partial_{\mathbf{1}\left(n\right)}\mathcal{H}_{\mathbf{1}\left(i\right)\mathbf{1}\left(j\right)}+\frac{1}{8}\mathcal{H}^{\mathbf{1}\left(m\right)\mathbf{1}\left(n\right)}\partial_{\mathbf{1}\left(m\right)}\mathcal{H}^{\mathbf{1}\left(i\right)\mathbf{2}\left(j\right)}\partial_{\mathbf{1}\left(n\right)}\mathcal{H}_{\mathbf{1}\left(i\right)\mathbf{2}\left(j\right)}\nonumber \\
 &  & +\frac{1}{8}\mathcal{H}^{\mathbf{1}\left(m\right)\mathbf{1}\left(n\right)}\partial_{\mathbf{1}\left(m\right)}\mathcal{H}^{\mathbf{2}\left(i\right)\mathbf{1}\left(j\right)}\partial_{\mathbf{1}\left(n\right)}\mathcal{H}_{\mathbf{2}\left(i\right)\mathbf{1}\left(j\right)}+\frac{1}{8}\mathcal{H}^{\mathbf{1}\left(m\right)\mathbf{1}\left(n\right)}\partial_{\mathbf{1}\left(m\right)}\mathcal{H}^{\mathbf{2}\left(i\right)\mathbf{2}\left(j\right)}\partial_{\mathbf{1}\left(n\right)}\mathcal{H}_{\mathbf{2}\left(i\right)\mathbf{2}\left(j\right)}\nonumber \\
 &  & +\frac{1}{8}\mathcal{H}^{\mathbf{1}\left(m\right)\mathbf{2}\left(n\right)}\partial_{\mathbf{1}\left(m\right)}\mathcal{H}^{\mathbf{1}\left(i\right)\mathbf{1}\left(j\right)}\partial_{\mathbf{2}\left(n\right)}\mathcal{H}_{\mathbf{1}\left(i\right)\mathbf{1}\left(j\right)}+\frac{1}{8}\mathcal{H}^{\mathbf{1}\left(m\right)\mathbf{2}\left(n\right)}\partial_{\mathbf{1}\left(m\right)}\mathcal{H}^{\mathbf{1}\left(i\right)\mathbf{2}\left(j\right)}\partial_{\mathbf{2}\left(n\right)}\mathcal{H}_{\mathbf{1}\left(i\right)\mathbf{2}\left(j\right)}\nonumber \\
 &  & +\frac{1}{8}\mathcal{H}^{\mathbf{1}\left(m\right)\mathbf{2}\left(n\right)}\partial_{\mathbf{1}\left(m\right)}\mathcal{H}^{\mathbf{2}\left(i\right)\mathbf{1}\left(j\right)}\partial_{\mathbf{2}\left(n\right)}\mathcal{H}_{\mathbf{2}\left(i\right)\mathbf{1}\left(j\right)}+\frac{1}{8}\mathcal{H}^{\mathbf{1}\left(m\right)\mathbf{2}\left(n\right)}\partial_{\mathbf{1}\left(m\right)}\mathcal{H}^{\mathbf{2}\left(i\right)\mathbf{2}\left(j\right)}\partial_{\mathbf{2}\left(n\right)}\mathcal{H}_{\mathbf{2}\left(i\right)\mathbf{2}\left(j\right)}\nonumber \\
 &  & +\frac{1}{8}\mathcal{H}^{\mathbf{2}\left(m\right)\mathbf{1}\left(n\right)}\partial_{\mathbf{2}\left(m\right)}\mathcal{H}^{\mathbf{1}\left(i\right)\mathbf{1}\left(j\right)}\partial_{\mathbf{1}\left(n\right)}\mathcal{H}_{\mathbf{1}\left(i\right)\mathbf{1}\left(j\right)}+\frac{1}{8}\mathcal{H}^{\mathbf{2}\left(m\right)\mathbf{1}\left(n\right)}\partial_{\mathbf{2}\left(m\right)}\mathcal{H}^{\mathbf{1}\left(i\right)\mathbf{2}\left(j\right)}\partial_{\mathbf{1}\left(n\right)}\mathcal{H}_{\mathbf{1}\left(i\right)\mathbf{2}\left(j\right)}\nonumber \\
 &  & +\frac{1}{8}\mathcal{H}^{\mathbf{2}\left(m\right)\mathbf{1}\left(n\right)}\partial_{\mathbf{2}\left(m\right)}\mathcal{H}^{\mathbf{2}\left(i\right)\mathbf{1}\left(j\right)}\partial_{\mathbf{1}\left(n\right)}\mathcal{H}_{\mathbf{2}\left(i\right)\mathbf{1}\left(j\right)}+\frac{1}{8}\mathcal{H}^{\mathbf{2}\left(m\right)\mathbf{1}\left(n\right)}\partial_{\mathbf{2}\left(m\right)}\mathcal{H}^{\mathbf{2}\left(i\right)\mathbf{2}\left(j\right)}\partial_{\mathbf{1}\left(n\right)}\mathcal{H}_{\mathbf{2}\left(i\right)\mathbf{2}\left(j\right)}\nonumber \\
 &  & +\frac{1}{8}\mathcal{H}^{\mathbf{2}\left(m\right)\mathbf{2}\left(n\right)}\partial_{\mathbf{2}\left(m\right)}\mathcal{H}^{\mathbf{1}\left(i\right)\mathbf{1}\left(j\right)}\partial_{\mathbf{2}\left(n\right)}\mathcal{H}_{\mathbf{1}\left(i\right)\mathbf{1}\left(j\right)}+\frac{1}{8}\mathcal{H}^{\mathbf{2}\left(m\right)\mathbf{2}\left(n\right)}\partial_{\mathbf{2}\left(m\right)}\mathcal{H}^{\mathbf{1}\left(i\right)\mathbf{2}\left(j\right)}\partial_{\mathbf{2}\left(n\right)}\mathcal{H}_{\mathbf{1}\left(i\right)\mathbf{2}\left(j\right)}\nonumber \\
 &  & +\frac{1}{8}\mathcal{H}^{\mathbf{2}\left(m\right)\mathbf{2}\left(n\right)}\partial_{\mathbf{2}\left(m\right)}\mathcal{H}^{\mathbf{2}\left(i\right)\mathbf{1}\left(j\right)}\partial_{\mathbf{2}\left(n\right)}\mathcal{H}_{\mathbf{2}\left(i\right)\mathbf{1}\left(j\right)}+\frac{1}{8}\mathcal{H}^{\mathbf{2}\left(m\right)\mathbf{2}\left(n\right)}\partial_{\mathbf{2}\left(m\right)}\mathcal{H}^{\mathbf{2}\left(i\right)\mathbf{2}\left(j\right)}\partial_{\mathbf{2}\left(n\right)}\mathcal{H}_{\mathbf{2}\left(i\right)\mathbf{2}\left(j\right)}\nonumber \\
\nonumber \\
 &  & -\frac{1}{2}\mathcal{H}^{\mathbf{1}\left(m\right)\mathbf{1}\left(n\right)}\partial_{\mathbf{1}\left(m\right)}\mathcal{H}^{\mathbf{1}\left(i\right)\mathbf{1}\left(j\right)}\partial_{\mathbf{1}\left(i\right)}\mathcal{H}_{\mathbf{1}\left(n\right)\mathbf{1}\left(j\right)}-\frac{1}{2}\mathcal{H}^{\mathbf{1}\left(m\right)\mathbf{1}\left(n\right)}\partial_{\mathbf{1}\left(m\right)}\mathcal{H}^{\mathbf{1}\left(i\right)\mathbf{2}\left(j\right)}\partial_{\mathbf{1}\left(i\right)}\mathcal{H}_{\mathbf{1}\left(n\right)\mathbf{2}\left(j\right)}\nonumber \\
 &  & -\frac{1}{2}\mathcal{H}^{\mathbf{1}\left(m\right)\mathbf{1}\left(n\right)}\partial_{\mathbf{1}\left(m\right)}\mathcal{H}^{\mathbf{2}\left(i\right)\mathbf{1}\left(j\right)}\partial_{\mathbf{2}\left(i\right)}\mathcal{H}_{\mathbf{1}\left(n\right)\mathbf{1}\left(j\right)}-\frac{1}{2}\mathcal{H}^{\mathbf{1}\left(m\right)\mathbf{1}\left(n\right)}\partial_{\mathbf{1}\left(m\right)}\mathcal{H}^{\mathbf{2}\left(i\right)\mathbf{2}\left(j\right)}\partial_{\mathbf{2}\left(i\right)}\mathcal{H}_{\mathbf{1}\left(n\right)\mathbf{2}\left(j\right)}\nonumber \\
 &  & -\frac{1}{2}\mathcal{H}^{\mathbf{1}\left(m\right)\mathbf{2}\left(n\right)}\partial_{\mathbf{1}\left(m\right)}\mathcal{H}^{\mathbf{1}\left(i\right)\mathbf{1}\left(j\right)}\partial_{\mathbf{1}\left(i\right)}\mathcal{H}_{\mathbf{2}\left(n\right)\mathbf{1}\left(j\right)}-\frac{1}{2}\mathcal{H}^{\mathbf{1}\left(m\right)\mathbf{2}\left(n\right)}\partial_{\mathbf{1}\left(m\right)}\mathcal{H}^{\mathbf{1}\left(i\right)\mathbf{2}\left(j\right)}\partial_{\mathbf{1}\left(i\right)}\mathcal{H}_{\mathbf{2}\left(n\right)\mathbf{2}\left(j\right)}\nonumber \\
 &  & -\frac{1}{2}\mathcal{H}^{\mathbf{1}\left(m\right)\mathbf{2}\left(n\right)}\partial_{\mathbf{1}\left(m\right)}\mathcal{H}^{\mathbf{2}\left(i\right)\mathbf{1}\left(j\right)}\partial_{\mathbf{2}\left(i\right)}\mathcal{H}_{\mathbf{2}\left(n\right)\mathbf{1}\left(j\right)}-\frac{1}{2}\mathcal{H}^{\mathbf{1}\left(m\right)\mathbf{2}\left(n\right)}\partial_{\mathbf{1}\left(m\right)}\mathcal{H}^{\mathbf{2}\left(i\right)\mathbf{2}\left(j\right)}\partial_{\mathbf{2}\left(i\right)}\mathcal{H}_{\mathbf{2}\left(n\right)\mathbf{2}\left(j\right)}\nonumber \\
 &  & -\frac{1}{2}\mathcal{H}^{\mathbf{2}\left(m\right)\mathbf{1}\left(n\right)}\partial_{\mathbf{2}\left(m\right)}\mathcal{H}^{\mathbf{1}\left(i\right)\mathbf{1}\left(j\right)}\partial_{\mathbf{1}\left(i\right)}\mathcal{H}_{\mathbf{1}\left(n\right)\mathbf{1}\left(j\right)}-\frac{1}{2}\mathcal{H}^{\mathbf{2}\left(m\right)\mathbf{1}\left(n\right)}\partial_{\mathbf{2}\left(m\right)}\mathcal{H}^{\mathbf{1}\left(i\right)\mathbf{2}\left(j\right)}\partial_{\mathbf{1}\left(i\right)}\mathcal{H}_{\mathbf{1}\left(n\right)\mathbf{2}\left(j\right)}\nonumber \\
 &  & -\frac{1}{2}\mathcal{H}^{\mathbf{2}\left(m\right)\mathbf{1}\left(n\right)}\partial_{\mathbf{2}\left(m\right)}\mathcal{H}^{\mathbf{2}\left(i\right)\mathbf{1}\left(j\right)}\partial_{\mathbf{2}\left(i\right)}\mathcal{H}_{\mathbf{1}\left(n\right)\mathbf{1}\left(j\right)}-\frac{1}{2}\mathcal{H}^{\mathbf{2}\left(m\right)\mathbf{1}\left(n\right)}\partial_{\mathbf{2}\left(m\right)}\mathcal{H}^{\mathbf{2}\left(i\right)\mathbf{2}\left(j\right)}\partial_{\mathbf{2}\left(i\right)}\mathcal{H}_{\mathbf{1}\left(n\right)\mathbf{2}\left(j\right)}\nonumber \\
 &  & -\frac{1}{2}\mathcal{H}^{\mathbf{2}\left(m\right)\mathbf{2}\left(n\right)}\partial_{\mathbf{2}\left(m\right)}\mathcal{H}^{\mathbf{1}\left(i\right)\mathbf{1}\left(j\right)}\partial_{\mathbf{1}\left(i\right)}\mathcal{H}_{\mathbf{2}\left(n\right)\mathbf{1}\left(j\right)}-\frac{1}{2}\mathcal{H}^{\mathbf{2}\left(m\right)\mathbf{2}\left(n\right)}\partial_{\mathbf{2}\left(m\right)}\mathcal{H}^{\mathbf{1}\left(i\right)\mathbf{2}\left(j\right)}\partial_{\mathbf{1}\left(i\right)}\mathcal{H}_{\mathbf{2}\left(n\right)\mathbf{2}\left(j\right)}\nonumber \\
 &  & -\frac{1}{2}\mathcal{H}^{\mathbf{2}\left(m\right)\mathbf{2}\left(n\right)}\partial_{\mathbf{2}\left(m\right)}\mathcal{H}^{\mathbf{2}\left(i\right)\mathbf{1}\left(j\right)}\partial_{\mathbf{2}\left(i\right)}\mathcal{H}_{\mathbf{2}\left(n\right)\mathbf{1}\left(j\right)}-\frac{1}{2}\mathcal{H}^{\mathbf{2}\left(m\right)\mathbf{2}\left(n\right)}\partial_{\mathbf{2}\left(m\right)}\mathcal{H}^{\mathbf{2}\left(i\right)\mathbf{2}\left(j\right)}\partial_{\mathbf{2}\left(i\right)}\mathcal{H}_{\mathbf{2}\left(n\right)\mathbf{2}\left(j\right)}\nonumber \\
\nonumber \\
 &  & -\partial_{\mathbf{1}\left(i\right)}\partial_{\mathbf{1}\left(j\right)}\mathcal{H}^{\mathbf{1}\left(i\right)\mathbf{1}\left(j\right)}-\partial_{\mathbf{1}\left(i\right)}\partial_{\mathbf{2}\left(j\right)}\mathcal{H}^{\mathbf{1}\left(i\right)\mathbf{2}\left(j\right)}-\partial_{\mathbf{2}\left(i\right)}\partial_{\mathbf{1}\left(j\right)}\mathcal{H}^{\mathbf{2}\left(i\right)\mathbf{1}\left(j\right)}-\partial_{\mathbf{2}\left(i\right)}\partial_{\mathbf{2}\left(j\right)}\mathcal{H}^{\mathbf{2}\left(i\right)\mathbf{2}\left(j\right)}\nonumber \\
\nonumber \\
 &  & -4\mathcal{H}^{\mathbf{1}\left(i\right)\mathbf{1}\left(j\right)}\partial_{\mathbf{1}\left(i\right)}d\partial_{\mathbf{1}\left(j\right)}d-4\mathcal{H}^{\mathbf{1}\left(i\right)\mathbf{2}\left(j\right)}\partial_{\mathbf{1}\left(i\right)}d\partial_{\mathbf{2}\left(j\right)}d-4\mathcal{H}^{\mathbf{2}\left(i\right)\mathbf{1}\left(j\right)}\partial_{\mathbf{2}\left(i\right)}d\partial_{\mathbf{1}\left(j\right)}d-4\mathcal{H}^{\mathbf{2}\left(i\right)\mathbf{2}\left(j\right)}\partial_{\mathbf{2}\left(i\right)}d\partial_{\mathbf{2}\left(j\right)}d\nonumber \\
\nonumber \\
 &  & 4\partial_{\mathbf{1}\left(i\right)}\mathcal{H}^{\mathbf{1}\left(i\right)\mathbf{1}\left(j\right)}\partial_{\mathbf{1}\left(j\right)}d+4\partial_{\mathbf{1}\left(i\right)}\mathcal{H}^{\mathbf{1}\left(i\right)\mathbf{2}\left(j\right)}\partial_{\mathbf{2}\left(j\right)}d+4\partial_{\mathbf{2}\left(i\right)}\mathcal{H}^{\mathbf{2}\left(i\right)\mathbf{1}\left(j\right)}\partial_{\mathbf{1}\left(j\right)}d+4\partial_{\mathbf{2}\left(i\right)}\mathcal{H}^{\mathbf{2}\left(i\right)\mathbf{2}\left(j\right)}\partial_{\mathbf{2}\left(j\right)}d\nonumber \\
\nonumber \\
 &  & 4\mathcal{H}^{\mathbf{1}\left(i\right)\mathbf{1}\left(j\right)}\partial_{\mathbf{1}\left(i\right)}\partial_{\mathbf{1}\left(j\right)}d+4\mathcal{H}^{\mathbf{1}\left(i\right)\mathbf{2}\left(j\right)}\partial_{\mathbf{1}\left(i\right)}\partial_{\mathbf{2}\left(j\right)}d+4\mathcal{H}^{\mathbf{2}\left(i\right)\mathbf{1}\left(j\right)}\partial_{\mathbf{2}\left(i\right)}\partial_{\mathbf{1}\left(j\right)}d+4\mathcal{H}^{\mathbf{2}\left(i\right)\mathbf{2}\left(j\right)}\partial_{\mathbf{2}\left(i\right)}\partial_{\mathbf{2}\left(j\right)}d.
\end{eqnarray}

\noindent Substituting the metric, one gets the result

\begin{equation}
3\left(\frac{\dot{\tilde{a}}^{2}}{\tilde{a}^{2}}\right)+3\left(\frac{\dot{a}^{2}}{a^{2}}\right)+4\dot{\tilde{d}}^{2}+4\dot{d}^{2}-4\ddot{\tilde{d}}-4\ddot{d}=0.
\end{equation}

\noindent We define the Hubble parameters as

\begin{equation}
H=\frac{\partial_{t}a}{a},\qquad\tilde{H}=\frac{\tilde{\partial}_{t}a}{a}.
\end{equation}

\noindent The EOM of the dilaton becomes

\begin{equation}
\left(3H^{2}-4\ddot{d}+4\dot{d}^{2}\right)+\left(3\tilde{H}^{2}-4\ddot{\tilde{d}}+4\dot{\tilde{d}}^{2}\right)=0.
\end{equation}

\subsection{Equation of motion of the generalized metric}

Recall the definition of the Ricci-like tensor

\begin{equation}
\mathcal{R}_{MN}=\mathcal{K}_{MN}-S_{\quad M}^{P}\mathcal{K}_{PQ}S_{\quad N}^{Q}=0.
\end{equation}

\noindent To make calculations simpler, we separate this equation into
two parts

\noindent
\begin{equation}
\mathcal{K}_{MN}=\star\mathcal{K}_{MN}+*\mathcal{K}_{MN},
\end{equation}

\noindent where the first part only includes a generalized metric

\noindent
\begin{eqnarray}
\star\mathcal{K}_{MN} & \equiv & \frac{1}{8}\partial_{M}\mathcal{H}^{KL}\partial_{N}\mathcal{H}_{KL}-\frac{1}{4}\partial_{L}\left(\mathcal{H}^{LK}\partial_{K}\mathcal{H}_{MN}\right)\nonumber \\
 &  & -\frac{1}{2}\partial_{\left(N\right.}\mathcal{H}^{KL}\partial_{L}\mathcal{H}_{\left.M\right)K}+\frac{1}{2}\partial_{L}\left(\mathcal{H}^{KL}\partial_{\left(N\right.}\mathcal{H}_{\left.M\right)K}+\mathcal{H}_{\quad\left(M\right.}^{K}\partial_{K}\mathcal{H}_{\quad\left.N\right)}^{L}\right),
\end{eqnarray}

\noindent and the second includes the dilaton

\noindent
\begin{equation}
*\mathcal{K}_{MN}\equiv\frac{1}{2}\partial_{L}d\left(\mathcal{H}^{LK}\partial_{K}\mathcal{H}_{MN}\right)+2\partial_{M}\partial_{N}d-\partial_{L}d\left(\mathcal{H}^{KL}\partial_{\left(N\right.}\mathcal{H}_{\left.M\right)K}+\mathcal{H}_{\quad\left(M\right.}^{K}\partial_{K}\mathcal{H}_{\quad\left.N\right)}^{L}\right).
\end{equation}

\noindent Therefore, the Ricci-like tensor can be rewritten as

\noindent
\begin{equation}
\mathcal{R}_{MN}=\star\mathcal{R}_{MN}+*\mathcal{R}_{MN}=0.
\end{equation}

\subsubsection{Calculation of $\star\mathcal{R}_{MN}$}

Expanding every term in $\star\mathcal{R}_{MN}$, we have



\noindent Using the calculation rules to simplify the equations, we get

%

\noindent Our aim is to calculate the components of $\star\mathcal{R}_{MN}$,
which can be expressed as

\noindent
\begin{equation}
\star\mathcal{R}_{MN}=\left(%

\noindent When $p=q=1$, we find

\noindent
\begin{eqnarray}
 &  & \star\mathcal{R}_{\mathbf{2}\left(t\right)\mathbf{2}\left(t\right)}\nonumber \\
 & = & \frac{1}{8}\partial_{t}\mathcal{G}_{ii}\partial_{t}\mathcal{G}^{ii}+\frac{1}{8}\partial_{t}\mathcal{B}_{i}^{\; j}\partial_{t}\mathcal{B}_{\; j}^{i}+\frac{1}{8}\partial_{t}\mathcal{C}_{\; j}^{i}\partial_{t}\mathcal{C}_{i}^{\; j}+\frac{1}{8}\partial_{t}\mathcal{D}^{ii}\partial_{t}\mathcal{D}_{ii}\nonumber \\
 &  & -\frac{1}{8}\mathcal{G}_{tt}\tilde{\partial}^{t}\mathcal{G}_{kk}\tilde{\partial}^{t}\mathcal{G}^{kk}\mathcal{G}_{tt}-\frac{1}{8}\mathcal{G}_{tt}\tilde{\partial}^{t}\mathcal{B}_{k}^{\; l}\tilde{\partial}^{t}\mathcal{B}_{\; l}^{k}\tilde{\mathcal{G}}_{tt}-\frac{1}{8}\mathcal{G}_{tt}\tilde{\partial}^{t}\mathcal{C}_{\; l}^{k}\tilde{\partial}^{t}\mathcal{C}_{k}^{\; l}\mathcal{G}_{tt}-\frac{1}{8}\mathcal{G}_{tt}\tilde{\partial}^{t}\mathcal{D}^{kk}\tilde{\partial}^{t}\mathcal{D}_{kk}\mathcal{G}_{tt}\nonumber \\
 & = & -3\frac{\dot{a}^{2}}{a^{2}}+3\frac{\dot{\tilde{a}}^{2}}{\tilde{a}^{2}}.
\end{eqnarray}

\noindent When $p=q=2$, we find

\begin{eqnarray}
 &  & \star\mathcal{R}_{\mathbf{2}\left(2\right)\mathbf{2}\left(2\right)}\nonumber \\
 & = & -\frac{1}{4}\partial_{t}\left(\mathcal{D}^{tt}\partial_{t}\mathcal{D}_{22}\right)+\frac{1}{4}\mathcal{G}_{22}\partial_{t}\left(\mathcal{D}^{tt}\partial_{t}\mathcal{G}^{22}\right)\mathcal{G}_{22}\nonumber \\
 &  & -\frac{1}{4}\tilde{\partial}^{t}\left(\mathcal{G}_{tt}\tilde{\partial}^{t}\mathcal{D}_{22}\right)+\frac{1}{4}\mathcal{G}_{22}\tilde{\partial}^{t}\left(\mathcal{G}_{tt}\tilde{\partial}^{t}\mathcal{G}^{22}\right)\mathcal{G}_{22}\nonumber \\
 & = & -\dot{a}^{2}+a\ddot{a}+\tilde{a}^{-4}\left(-\tilde{a}\ddot{\tilde{a}}+\dot{\tilde{a}}^{2}\right).
\end{eqnarray}

\noindent When $p=q=3,4$, we find

\begin{eqnarray}
 &  & \star\mathcal{R}_{\mathbf{2}\left(3,4\right)\mathbf{2}\left(3,4\right)}\nonumber \\
 & = & -\frac{1}{4}\partial_{t}\left(\mathcal{D}^{tt}\partial_{t}\mathcal{D}_{33}\right)+\frac{1}{4}\mathcal{G}_{33}\partial_{t}\left(\mathcal{D}^{tt}\partial_{t}\mathcal{G}^{33}\right)\mathcal{G}_{33}+\frac{1}{4}\mathcal{G}_{33}\partial_{t}\left(\mathcal{D}^{tt}\partial_{t}\mathcal{B}_{\;4}^{3}\right)\mathcal{C}_{\;3}^{4}\nonumber \\
 &  & +\frac{1}{4}\mathcal{C}_{\;3}^{4}\partial_{t}\left(\mathcal{D}^{tt}\partial_{t}\mathcal{C}_{4}^{\;3}\right)\mathcal{G}_{33}+\frac{1}{4}\mathcal{C}_{\;3}^{4}\partial_{t}\left(\mathcal{D}^{tt}\partial_{t}\mathcal{D}_{44}\right)\mathcal{C}_{\;3}^{4}\nonumber \\
 &  & -\frac{1}{4}\tilde{\partial}^{t}\left(\mathcal{G}_{tt}\tilde{\partial}^{t}\mathcal{D}_{33}\right)+\frac{1}{4}\mathcal{G}_{33}\tilde{\partial}^{t}\left(\mathcal{G}_{tt}\tilde{\partial}^{t}\mathcal{G}^{33}\right)\mathcal{G}_{33}+\frac{1}{4}\mathcal{G}_{33}\tilde{\partial}^{t}\left(\mathcal{D}^{tt}\tilde{\partial}^{t}\mathcal{B}_{\;4}^{3}\right)\mathcal{C}_{\;3}^{4}\nonumber \\
 &  & +\frac{1}{4}\mathcal{C}_{\;3}^{4}\tilde{\partial}^{t}\left(\mathcal{G}_{tt}\tilde{\partial}^{t}\mathcal{C}_{4}^{\;3}\right)\mathcal{G}_{33}+\frac{1}{4}\mathcal{C}_{\;3}^{4}\tilde{\partial}^{t}\left(\mathcal{G}_{tt}\tilde{\partial}^{t}\mathcal{D}_{44}\right)\mathcal{C}_{\;3}^{4}\nonumber \\
 & = & \left(1-a^{-4}b^{2}\right)\left(-\dot{a}^{2}+a\ddot{a}\right)+\left(\tilde{a}{}^{-4}-b^{2}\right)\left(\dot{\tilde{a}}^{2}-\tilde{a}\ddot{\tilde{a}}\right).
\end{eqnarray}

\noindent Finally, we have

\begin{eqnarray}
\star\mathcal{R}_{\mathbf{2}\left(t\right)\mathbf{2}\left(t\right)} & = & -\left[3\frac{\dot{a}^{2}}{a^{2}}\right]+\left[3\frac{\dot{\tilde{a}}^{2}}{\tilde{a}^{2}}\right],\nonumber \\
\star\mathcal{R}_{\mathbf{2}\left(2\right)\mathbf{2}\left(2\right)} & = & -\left[\dot{a}^{2}-a\ddot{a}\right]+\tilde{a}^{-4}\left[\dot{\tilde{a}}^{2}-\tilde{a}\ddot{\tilde{a}}\right],\nonumber \\
\star\mathcal{R}_{\mathbf{2}\left(3,4\right)\mathbf{2}\left(3,4\right)} & = & -\left(1-a^{-4}b^{2}\right)\left[\dot{a}^{2}-a\ddot{a}\right]+\left(\tilde{a}{}^{-4}-b^{2}\right)\left[\dot{\tilde{a}}^{2}-\tilde{a}\ddot{\tilde{a}}\right].
\end{eqnarray}

\subsubsection{Calculation of $*\mathcal{R}_{MN}$}



\noindent Using our notations, we find

\begin{eqnarray}
 &  & *\mathcal{R}_{MN}\nonumber \\
 & = & *\mathcal{K}_{MN}-S_{\quad M}^{P}*\mathcal{K}_{PQ}S_{\quad N}^{Q}\nonumber \\
 & = & \frac{1}{2}\tilde{\partial}^{i}d\left(\mathcal{G}_{ij}\tilde{\partial}^{j}\mathcal{H}_{MN}\right)+\frac{1}{2}\tilde{\partial}^{i}d\left(\mathcal{B}_{i}^{\; j}\partial_{j}\mathcal{H}_{MN}\right)+\frac{1}{2}\partial_{i}d\left(\mathcal{C}_{\; j}^{i}\tilde{\partial}^{j}\mathcal{H}_{MN}\right)+\frac{1}{2}\partial_{i}d\left(\mathcal{D}^{ij}\partial_{j}\mathcal{H}_{MN}\right)\nonumber \\
 &  & +2\partial_{M}\partial_{N}d\nonumber \\
 &  & -\tilde{\partial}^{i}d\left(\mathcal{G}_{ji}\partial_{\left(N\right.}\mathcal{H}_{\left.M\right)\mathbf{1}\left(j\right)}\right)-\tilde{\partial}^{i}d\left(\mathcal{C}_{\; i}^{j}\partial_{\left(N\right.}\mathcal{H}_{\left.M\right)\mathbf{2}\left(j\right)}\right)-\partial_{i}d\left(\mathcal{B}_{j}^{\; i}\partial_{\left(N\right.}\mathcal{H}_{\left.M\right)\mathbf{1}\left(j\right)}\right)-\partial_{i}d\left(\mathcal{D}^{ji}\partial_{\left(N\right.}\mathcal{H}_{\left.M\right)\mathbf{2}\left(j\right)}\right)\nonumber \\
 &  & -\tilde{\partial}^{i}d\left(\mathcal{H}_{\quad\left(M\right.}^{\mathbf{1}\left(j\right)}\tilde{\partial}^{j}\mathcal{H}_{\quad\left.N\right)}^{\mathbf{1}\left(i\right)}\right)-\tilde{\partial}^{i}d\left(\mathcal{H}_{\quad\left(M\right.}^{\mathbf{2}\left(j\right)}\partial_{j}\mathcal{H}_{\quad\left.N\right)}^{\mathbf{1}\left(i\right)}\right)-\partial_{i}d\left(\mathcal{H}_{\quad\left(M\right.}^{\mathbf{1}\left(j\right)}\tilde{\partial}^{j}\mathcal{H}_{\quad\left.N\right)}^{\mathbf{2}\left(i\right)}\right)-\partial_{i}d\left(\mathcal{H}_{\quad\left(M\right.}^{\mathbf{2}\left(j\right)}\partial_{j}\mathcal{H}_{\quad\left.N\right)}^{\mathbf{2}\left(i\right)}\right)\nonumber \\
\nonumber \\
 &  & -\frac{1}{2}S_{\quad M}^{\mathbf{1}\left(i\right)}\tilde{\partial}^{l}d\left(\mathcal{G}_{lk}\tilde{\partial}^{k}\mathcal{G}^{ij}\right)S_{\quad N}^{\mathbf{1}\left(j\right)}-\frac{1}{2}S_{\quad M}^{\mathbf{1}\left(i\right)}\tilde{\partial}^{l}d\left(\mathcal{B}_{l}^{\; k}\partial_{k}\mathcal{G}^{ij}\right)S_{\quad N}^{\mathbf{1}\left(j\right)}-\frac{1}{2}S_{\quad M}^{\mathbf{1}\left(i\right)}\partial_{l}d\left(\mathcal{C}_{\; k}^{l}\tilde{\partial}^{k}\mathcal{G}^{ij}\right)S_{\quad N}^{\mathbf{1}\left(j\right)}\nonumber \\
 &  & -\frac{1}{2}S_{\quad M}^{\mathbf{1}\left(i\right)}\tilde{\partial}^{l}d\left(\mathcal{G}_{lk}\tilde{\partial}^{k}\mathcal{B}_{\; j}^{i}\right)S_{\quad N}^{\mathbf{2}\left(j\right)}-\frac{1}{2}S_{\quad M}^{\mathbf{1}\left(i\right)}\tilde{\partial}^{l}d\left(\mathcal{B}_{l}^{\; k}\partial_{k}\mathcal{B}_{\; j}^{i}\right)S_{\quad N}^{\mathbf{2}\left(j\right)}-\frac{1}{2}S_{\quad M}^{\mathbf{1}\left(i\right)}\partial_{l}d\left(\mathcal{C}_{\; k}^{l}\tilde{\partial}^{k}\mathcal{B}_{\; j}^{i}\right)S_{\quad N}^{\mathbf{2}\left(j\right)}\nonumber \\
 &  & -\frac{1}{2}S_{\quad M}^{\mathbf{2}\left(i\right)}\tilde{\partial}^{l}d\left(\mathcal{G}_{lk}\tilde{\partial}^{k}\mathcal{C}_{i}^{\; j}\right)S_{\quad N}^{\mathbf{1}\left(j\right)}-\frac{1}{2}S_{\quad M}^{\mathbf{2}\left(i\right)}\tilde{\partial}^{l}d\left(\mathcal{B}_{l}^{\; k}\partial_{k}\mathcal{C}_{i}^{\; j}\right)S_{\quad N}^{\mathbf{1}\left(j\right)}-\frac{1}{2}S_{\quad M}^{\mathbf{2}\left(i\right)}\partial_{l}d\left(\mathcal{C}_{\; k}^{l}\tilde{\partial}^{k}\mathcal{C}_{i}^{\; j}\right)S_{\quad N}^{\mathbf{1}\left(j\right)}\nonumber \\
 &  & -\frac{1}{2}S_{\quad M}^{\mathbf{2}\left(i\right)}\tilde{\partial}^{l}d\left(\mathcal{G}_{lk}\tilde{\partial}^{k}\mathcal{D}_{ij}\right)S_{\quad N}^{\mathbf{2}\left(j\right)}-\frac{1}{2}S_{\quad M}^{\mathbf{2}\left(i\right)}\tilde{\partial}^{l}d\left(\mathcal{B}_{l}^{\; k}\partial_{k}\mathcal{D}_{ij}\right)S_{\quad N}^{\mathbf{2}\left(j\right)}-\frac{1}{2}S_{\quad M}^{\mathbf{2}\left(i\right)}\partial_{l}d\left(\mathcal{C}_{\; k}^{l}\tilde{\partial}^{k}\mathcal{D}_{ij}\right)S_{\quad N}^{\mathbf{2}\left(j\right)}\nonumber \\
 &  & -\frac{1}{2}S_{\quad M}^{\mathbf{1}\left(i\right)}\partial_{l}d\left(\mathcal{D}^{lk}\partial_{k}\mathcal{G}^{ij}\right)S_{\quad N}^{\mathbf{1}\left(j\right)}-\frac{1}{2}S_{\quad M}^{\mathbf{1}\left(i\right)}\partial_{l}d\left(\mathcal{D}^{lk}\partial_{k}\mathcal{B}_{\; j}^{i}\right)S_{\quad N}^{\mathbf{2}\left(j\right)}-\frac{1}{2}S_{\quad M}^{\mathbf{2}\left(i\right)}\partial_{l}d\left(\mathcal{D}^{lk}\partial_{k}\mathcal{C}_{i}^{\; j}\right)S_{\quad N}^{\mathbf{1}\left(j\right)}\nonumber \\
 &  & -\frac{1}{2}S_{\quad M}^{\mathbf{2}\left(i\right)}\partial_{l}d\left(\mathcal{D}^{lk}\partial_{k}\mathcal{D}_{ij}\right)S_{\quad N}^{\mathbf{2}\left(j\right)}\nonumber \\
\nonumber \\
 &  & -2S_{\quad M}^{\mathbf{1}\left(i\right)}\tilde{\partial}^{i}\tilde{\partial}^{j}dS_{\quad N}^{\mathbf{1}\left(j\right)}-2S_{\quad M}^{\mathbf{1}\left(i\right)}\tilde{\partial}^{i}\partial_{j}dS_{\quad N}^{\mathbf{2}\left(j\right)}-2S_{\quad M}^{\mathbf{2}\left(i\right)}\partial_{i}\tilde{\partial}^{j}dS_{\quad N}^{\mathbf{1}\left(j\right)}-2S_{\quad M}^{\mathbf{2}\left(i\right)}\partial_{i}\partial_{j}dS_{\quad N}^{\mathbf{2}\left(j\right)}\nonumber \\
\nonumber \\
 &  & +\frac{1}{2}S_{\quad M}^{\mathbf{1}\left(i\right)}\tilde{\partial}^{l}d\left(\mathcal{G}_{kl}\tilde{\partial}^{j}\mathcal{G}^{ik}\right)S_{\quad N}^{\mathbf{1}\left(j\right)}+\frac{1}{2}S_{\quad M}^{\mathbf{1}\left(i\right)}\tilde{\partial}^{l}d\left(\mathcal{C}_{\; l}^{k}\tilde{\partial}^{j}\mathcal{B}_{\; k}^{i}\right)S_{\quad N}^{\mathbf{1}\left(j\right)}+\frac{1}{2}S_{\quad M}^{\mathbf{1}\left(i\right)}\partial_{l}d\left(\mathcal{B}_{k}^{\; l}\tilde{\partial}^{j}\mathcal{G}^{ik}\right)S_{\quad N}^{\mathbf{1}\left(j\right)}\nonumber \\
 &  & +\frac{1}{2}S_{\quad M}^{\mathbf{1}\left(i\right)}\tilde{\partial}^{l}d\left(\mathcal{G}_{kl}\tilde{\partial}^{i}\mathcal{G}^{jk}\right)S_{\quad N}^{\mathbf{1}\left(j\right)}+\frac{1}{2}S_{\quad M}^{\mathbf{1}\left(i\right)}\tilde{\partial}^{l}d\left(\mathcal{C}_{\; l}^{k}\tilde{\partial}^{i}\mathcal{B}_{\; k}^{j}\right)S_{\quad N}^{\mathbf{1}\left(j\right)}+\frac{1}{2}S_{\quad M}^{\mathbf{1}\left(i\right)}\partial_{l}d\left(\mathcal{B}_{k}^{\; l}\tilde{\partial}^{i}\mathcal{G}^{jk}\right)S_{\quad N}^{\mathbf{1}\left(j\right)}\nonumber \\
 &  & +\frac{1}{2}S_{\quad M}^{\mathbf{1}\left(i\right)}\tilde{\partial}^{l}d\left(\mathcal{G}_{kl}\partial_{j}\mathcal{G}^{ik}\right)S_{\quad N}^{\mathbf{2}\left(j\right)}+\frac{1}{2}S_{\quad M}^{\mathbf{1}\left(i\right)}\tilde{\partial}^{l}d\left(\mathcal{C}_{\; l}^{k}\partial_{j}\mathcal{B}_{\; k}^{i}\right)S_{\quad N}^{\mathbf{2}\left(j\right)}+\frac{1}{2}S_{\quad M}^{\mathbf{1}\left(i\right)}\partial_{l}d\left(\mathcal{B}_{k}^{\; l}\partial_{j}\mathcal{G}^{ik}\right)S_{\quad N}^{\mathbf{2}\left(j\right)}\nonumber \\
 &  & +\frac{1}{2}S_{\quad M}^{\mathbf{1}\left(i\right)}\tilde{\partial}^{l}d\left(\mathcal{G}_{kl}\tilde{\partial}^{i}\mathcal{C}_{j}^{\; k}\right)S_{\quad N}^{\mathbf{2}\left(j\right)}+\frac{1}{2}S_{\quad M}^{\mathbf{1}\left(i\right)}\tilde{\partial}^{l}d\left(\mathcal{C}_{\; l}^{k}\tilde{\partial}^{i}\mathcal{D}_{jk}\right)S_{\quad N}^{\mathbf{2}\left(j\right)}+\frac{1}{2}S_{\quad M}^{\mathbf{1}\left(i\right)}\partial_{l}d\left(\mathcal{B}_{k}^{\; l}\tilde{\partial}^{i}\mathcal{C}_{j}^{\; k}\right)S_{\quad N}^{\mathbf{2}\left(j\right)}\nonumber \\
 &  & +\frac{1}{2}S_{\quad M}^{\mathbf{2}\left(i\right)}\tilde{\partial}^{l}d\left(\mathcal{G}_{kl}\tilde{\partial}^{j}\mathcal{C}_{i}^{\; k}\right)S_{\quad N}^{\mathbf{1}\left(j\right)}+\frac{1}{2}S_{\quad M}^{\mathbf{2}\left(i\right)}\tilde{\partial}^{l}d\left(\mathcal{C}_{\; l}^{k}\tilde{\partial}^{j}\mathcal{D}_{ik}\right)S_{\quad N}^{\mathbf{1}\left(j\right)}+\frac{1}{2}S_{\quad M}^{\mathbf{2}\left(i\right)}\partial_{l}d\left(\mathcal{B}_{k}^{\; l}\tilde{\partial}^{j}\mathcal{C}_{i}^{\; k}\right)S_{\quad N}^{\mathbf{1}\left(j\right)}\nonumber \\
 &  & +\frac{1}{2}S_{\quad M}^{\mathbf{2}\left(i\right)}\tilde{\partial}^{l}d\left(\mathcal{G}_{kl}\partial_{i}\mathcal{G}^{jk}\right)S_{\quad N}^{\mathbf{1}\left(j\right)}+\frac{1}{2}S_{\quad M}^{\mathbf{2}\left(i\right)}\tilde{\partial}^{l}d\left(\mathcal{C}_{\; l}^{k}\partial_{i}\mathcal{B}_{\; k}^{j}\right)S_{\quad N}^{\mathbf{1}\left(j\right)}+\frac{1}{2}S_{\quad M}^{\mathbf{2}\left(i\right)}\partial_{l}d\left(\mathcal{B}_{k}^{\; l}\partial_{i}\mathcal{G}^{jk}\right)S_{\quad N}^{\mathbf{1}\left(j\right)}\nonumber \\
 &  & +\frac{1}{2}S_{\quad M}^{\mathbf{2}\left(i\right)}\tilde{\partial}^{l}d\left(\mathcal{G}_{kl}\partial_{j}\mathcal{C}_{i}^{\; k}\right)S_{\quad N}^{\mathbf{2}\left(j\right)}+\frac{1}{2}S_{\quad M}^{\mathbf{2}\left(i\right)}\tilde{\partial}^{l}d\left(\mathcal{C}_{\; l}^{k}\partial_{j}\mathcal{D}_{ik}\right)S_{\quad N}^{\mathbf{2}\left(j\right)}+\frac{1}{2}S_{\quad M}^{\mathbf{2}\left(i\right)}\partial_{l}d\left(\mathcal{B}_{k}^{\; l}\partial_{j}\mathcal{C}_{i}^{\; k}\right)S_{\quad N}^{\mathbf{2}\left(j\right)}\nonumber \\
 &  & +\frac{1}{2}S_{\quad M}^{\mathbf{2}\left(i\right)}\tilde{\partial}^{l}d\left(\mathcal{G}_{kl}\partial_{i}\mathcal{C}_{j}^{\; k}\right)S_{\quad N}^{\mathbf{2}\left(j\right)}+\frac{1}{2}S_{\quad M}^{\mathbf{2}\left(i\right)}\tilde{\partial}^{l}d\left(\mathcal{C}_{\; l}^{k}\partial_{i}\mathcal{D}_{jk}\right)S_{\quad N}^{\mathbf{2}\left(j\right)}+\frac{1}{2}S_{\quad M}^{\mathbf{2}\left(i\right)}\partial_{l}d\left(\mathcal{B}_{k}^{\; l}\partial_{i}\mathcal{C}_{j}^{\; k}\right)S_{\quad N}^{\mathbf{2}\left(j\right)}\nonumber \\
 &  & +\frac{1}{2}S_{\quad M}^{\mathbf{1}\left(i\right)}\partial_{l}d\left(\mathcal{D}^{kl}\tilde{\partial}^{j}\mathcal{B}_{\; k}^{i}\right)S_{\quad N}^{\mathbf{1}\left(j\right)}+\frac{1}{2}S_{\quad M}^{\mathbf{1}\left(i\right)}\partial_{l}d\left(\mathcal{D}^{kl}\tilde{\partial}^{i}\mathcal{B}_{\; k}^{j}\right)S_{\quad N}^{\mathbf{1}\left(j\right)}+\frac{1}{2}S_{\quad M}^{\mathbf{1}\left(i\right)}\partial_{l}d\left(\mathcal{D}^{kl}\partial_{j}\mathcal{B}_{\; k}^{i}\right)S_{\quad N}^{\mathbf{2}\left(j\right)}\nonumber \\
 &  & +\frac{1}{2}S_{\quad M}^{\mathbf{1}\left(i\right)}\partial_{l}d\left(\mathcal{D}^{kl}\tilde{\partial}^{i}\mathcal{D}_{jk}\right)S_{\quad N}^{\mathbf{2}\left(j\right)}+\frac{1}{2}S_{\quad M}^{\mathbf{2}\left(i\right)}\partial_{l}d\left(\mathcal{D}^{kl}\tilde{\partial}^{j}\mathcal{D}_{ik}\right)S_{\quad N}^{\mathbf{1}\left(j\right)}+\frac{1}{2}S_{\quad M}^{\mathbf{2}\left(i\right)}\partial_{l}d\left(\mathcal{D}^{kl}\partial_{i}\mathcal{B}_{\; k}^{j}\right)S_{\quad N}^{\mathbf{1}\left(j\right)}\nonumber \\
 &  & +\frac{1}{2}S_{\quad M}^{\mathbf{2}\left(i\right)}\partial_{l}d\left(\mathcal{D}^{kl}\partial_{j}\mathcal{D}_{ik}\right)S_{\quad N}^{\mathbf{2}\left(j\right)}+\frac{1}{2}S_{\quad M}^{\mathbf{2}\left(i\right)}\partial_{l}d\left(\mathcal{D}^{kl}\partial_{i}\mathcal{D}_{jk}\right)S_{\quad N}^{\mathbf{2}\left(j\right)}\nonumber \\
\nonumber \\
 &  & +\frac{1}{2}S_{\quad M}^{\mathbf{1}\left(i\right)}\tilde{\partial}^{l}d\left(\mathcal{B}_{k}^{\; i}\tilde{\partial}^{k}\mathcal{B}_{l}^{\; j}\right)S_{\quad N}^{\mathbf{1}\left(j\right)}+\frac{1}{2}S_{\quad M}^{\mathbf{1}\left(i\right)}\tilde{\partial}^{l}d\left(\mathcal{D}^{ki}\partial_{k}\mathcal{B}_{l}^{\; j}\right)S_{\quad N}^{\mathbf{1}\left(j\right)}+\frac{1}{2}S_{\quad M}^{\mathbf{1}\left(i\right)}\partial_{l}d\left(\mathcal{B}_{k}^{\; i}\tilde{\partial}^{k}\mathcal{D}^{lj}\right)S_{\quad N}^{\mathbf{1}\left(j\right)}\nonumber \\
 &  & +\frac{1}{2}S_{\quad M}^{\mathbf{1}\left(i\right)}\tilde{\partial}^{l}d\left(\mathcal{B}_{k}^{\; j}\tilde{\partial}^{k}\mathcal{B}_{l}^{\; i}\right)S_{\quad N}^{\mathbf{1}\left(j\right)}+\frac{1}{2}S_{\quad M}^{\mathbf{1}\left(i\right)}\tilde{\partial}^{l}d\left(\mathcal{D}^{kj}\partial_{k}\mathcal{B}_{l}^{\; i}\right)S_{\quad N}^{\mathbf{1}\left(j\right)}+\frac{1}{2}S_{\quad M}^{\mathbf{1}\left(i\right)}\partial_{l}d\left(\mathcal{B}_{k}^{\; j}\tilde{\partial}^{k}\mathcal{D}^{li}\right)S_{\quad N}^{\mathbf{1}\left(j\right)}\nonumber \\
 &  & +\frac{1}{2}S_{\quad M}^{\mathbf{1}\left(i\right)}\tilde{\partial}^{l}d\left(\mathcal{B}_{k}^{\; i}\tilde{\partial}^{k}\mathcal{G}_{lj}\right)S_{\quad N}^{\mathbf{2}\left(j\right)}+\frac{1}{2}S_{\quad M}^{\mathbf{1}\left(i\right)}\tilde{\partial}^{l}d\left(\mathcal{D}^{ki}\partial_{k}\mathcal{G}_{lj}\right)S_{\quad N}^{\mathbf{2}\left(j\right)}+\frac{1}{2}S_{\quad M}^{\mathbf{1}\left(i\right)}\partial_{l}d\left(\mathcal{B}_{k}^{\; i}\tilde{\partial}^{k}\mathcal{C}_{\; j}^{l}\right)S_{\quad N}^{\mathbf{2}\left(j\right)}\nonumber \\
 &  & +\frac{1}{2}S_{\quad M}^{\mathbf{1}\left(i\right)}\tilde{\partial}^{l}d\left(\mathcal{G}_{kj}\tilde{\partial}^{k}\mathcal{B}_{l}^{\; i}\right)S_{\quad N}^{\mathbf{2}\left(j\right)}+\frac{1}{2}S_{\quad M}^{\mathbf{1}\left(i\right)}\tilde{\partial}^{l}d\left(\mathcal{C}_{\; j}^{k}\partial_{k}\mathcal{B}_{l}^{\; i}\right)S_{\quad N}^{\mathbf{2}\left(j\right)}+\frac{1}{2}S_{\quad M}^{\mathbf{1}\left(i\right)}\partial_{l}d\left(\mathcal{G}_{kj}\tilde{\partial}^{k}\mathcal{D}^{li}\right)S_{\quad N}^{\mathbf{2}\left(j\right)}\nonumber \\
 &  & +\frac{1}{2}S_{\quad M}^{\mathbf{2}\left(i\right)}\tilde{\partial}^{l}d\left(\mathcal{G}_{ki}\tilde{\partial}^{k}\mathcal{B}_{l}^{\; j}\right)S_{\quad N}^{\mathbf{1}\left(j\right)}+\frac{1}{2}S_{\quad M}^{\mathbf{2}\left(i\right)}\tilde{\partial}^{l}d\left(\mathcal{C}_{\; i}^{k}\partial_{k}\mathcal{B}_{l}^{\; j}\right)S_{\quad N}^{\mathbf{1}\left(j\right)}+\frac{1}{2}S_{\quad M}^{\mathbf{2}\left(i\right)}\partial_{l}d\left(\mathcal{G}_{ki}\tilde{\partial}^{k}\mathcal{D}^{lj}\right)S_{\quad N}^{\mathbf{1}\left(j\right)}\nonumber \\
 &  & +\frac{1}{2}S_{\quad M}^{\mathbf{2}\left(i\right)}\tilde{\partial}^{l}d\left(\mathcal{B}_{k}^{\; j}\tilde{\partial}^{k}\mathcal{G}_{li}\right)S_{\quad N}^{\mathbf{1}\left(j\right)}+\frac{1}{2}S_{\quad M}^{\mathbf{2}\left(i\right)}\tilde{\partial}^{l}d\left(\mathcal{D}^{kj}\partial_{k}\mathcal{G}_{li}\right)S_{\quad N}^{\mathbf{1}\left(j\right)}+\frac{1}{2}S_{\quad M}^{\mathbf{2}\left(i\right)}\partial_{l}d\left(\mathcal{B}_{k}^{\; j}\tilde{\partial}^{k}\mathcal{C}_{\; i}^{l}\right)S_{\quad N}^{\mathbf{1}\left(j\right)}\nonumber \\
 &  & +\frac{1}{2}S_{\quad M}^{\mathbf{2}\left(i\right)}\tilde{\partial}^{l}d\left(\mathcal{G}_{ki}\tilde{\partial}^{k}\mathcal{G}_{lj}\right)S_{\quad N}^{\mathbf{2}\left(j\right)}+\frac{1}{2}S_{\quad M}^{\mathbf{2}\left(i\right)}\tilde{\partial}^{l}d\left(\mathcal{C}_{\; i}^{k}\partial_{k}\mathcal{G}_{lj}\right)S_{\quad N}^{\mathbf{2}\left(j\right)}+\frac{1}{2}S_{\quad M}^{\mathbf{2}\left(i\right)}\partial_{l}d\left(\mathcal{G}_{ki}\tilde{\partial}^{k}\mathcal{C}_{\; j}^{l}\right)S_{\quad N}^{\mathbf{2}\left(j\right)}\nonumber \\
 &  & +\frac{1}{2}S_{\quad M}^{\mathbf{2}\left(i\right)}\tilde{\partial}^{l}d\left(\mathcal{G}_{kj}\tilde{\partial}^{k}\mathcal{G}_{li}\right)S_{\quad N}^{\mathbf{2}\left(j\right)}+\frac{1}{2}S_{\quad M}^{\mathbf{2}\left(i\right)}\tilde{\partial}^{l}d\left(\mathcal{C}_{\; j}^{k}\partial_{k}\mathcal{G}_{li}\right)S_{\quad N}^{\mathbf{2}\left(j\right)}+\frac{1}{2}S_{\quad M}^{\mathbf{2}\left(i\right)}\partial_{l}d\left(\mathcal{G}_{kj}\tilde{\partial}^{k}\mathcal{C}_{\; i}^{l}\right)S_{\quad N}^{\mathbf{2}\left(j\right)}\nonumber \\
 &  & +\frac{1}{2}S_{\quad M}^{\mathbf{1}\left(i\right)}\partial_{l}d\left(\mathcal{D}^{ki}\partial_{k}\mathcal{D}^{lj}\right)S_{\quad N}^{\mathbf{1}\left(j\right)}+\frac{1}{2}S_{\quad M}^{\mathbf{1}\left(i\right)}\partial_{l}d\left(\mathcal{D}^{kj}\partial_{k}\mathcal{D}^{li}\right)S_{\quad N}^{\mathbf{1}\left(j\right)}+\frac{1}{2}S_{\quad M}^{\mathbf{1}\left(i\right)}\partial_{l}d\left(\mathcal{D}^{ki}\partial_{k}\mathcal{C}_{\; j}^{l}\right)S_{\quad N}^{\mathbf{2}\left(j\right)}\nonumber \\
 &  & +\frac{1}{2}S_{\quad M}^{\mathbf{1}\left(i\right)}\partial_{l}d\left(\mathcal{C}_{\; j}^{k}\partial_{k}\mathcal{D}^{li}\right)S_{\quad N}^{\mathbf{2}\left(j\right)}+\frac{1}{2}S_{\quad M}^{\mathbf{2}\left(i\right)}\partial_{l}d\left(\mathcal{C}_{\; i}^{k}\partial_{k}\mathcal{D}^{lj}\right)S_{\quad N}^{\mathbf{1}\left(j\right)}+\frac{1}{2}S_{\quad M}^{\mathbf{2}\left(i\right)}\partial_{l}d\left(\mathcal{D}^{kj}\partial_{k}\mathcal{C}_{\; i}^{l}\right)S_{\quad N}^{\mathbf{1}\left(j\right)}\nonumber \\
 &  & +\frac{1}{2}S_{\quad M}^{\mathbf{2}\left(i\right)}\partial_{l}d\left(\mathcal{C}_{\; i}^{k}\partial_{k}\mathcal{C}_{\; j}^{l}\right)S_{\quad N}^{\mathbf{2}\left(j\right)}+\frac{1}{2}S_{\quad M}^{\mathbf{2}\left(i\right)}\partial_{l}d\left(\mathcal{C}_{\; j}^{k}\partial_{k}\mathcal{C}_{\; i}^{l}\right)S_{\quad N}^{\mathbf{2}\left(j\right)}.
\end{eqnarray}

\noindent $*\mathcal{R}_{MN}$ also can be expressed as

\begin{equation}
*\mathcal{R}_{MN}=\left(\begin{array}{cc}
*\mathcal{R}_{\mathbf{1}\mathbf{1}} & *\mathcal{R}_{\mathbf{1}\mathbf{2}}\\
*\mathcal{R}_{\mathbf{2}\mathbf{1}} & *\mathcal{R}_{\mathbf{2}\mathbf{2}}
\end{array}\right).
\end{equation}

\noindent When $p=q=1$, we get

\noindent
\begin{eqnarray}
*\mathcal{R}_{\mathbf{2}\left(t\right)\mathbf{2}\left(t\right)} & = & 2\partial_{t}\partial_{t}d-2\mathcal{G}_{tt}\tilde{\partial}^{t}\tilde{\partial}^{t}d\mathcal{G}_{tt}\nonumber \\
 & = & 2\ddot{d}-2\ddot{\tilde{d}}.
\end{eqnarray}

\noindent When $p=q=2$, the result is

\noindent
\begin{eqnarray}
 &  & *\mathcal{R}_{\mathbf{2}\left(2\right)\mathbf{2}\left(2\right)}\nonumber \\
 & = & \frac{1}{2}\tilde{\partial}^{t}d\left(\mathcal{G}_{tt}\tilde{\partial}^{t}\mathcal{D}_{22}\right)-\frac{1}{2}\mathcal{G}_{22}\tilde{\partial}^{t}d\left(\mathcal{G}_{tt}\tilde{\partial}^{t}\mathcal{G}^{22}\right)\mathcal{G}_{22}\nonumber \\
 &  & +\frac{1}{2}\partial_{t}d\left(\mathcal{D}^{tt}\partial_{t}\mathcal{D}_{22}\right)-\frac{1}{2}\mathcal{G}_{22}\partial_{t}d\left(\mathcal{D}^{tt}\partial_{t}\mathcal{G}^{22}\right)\mathcal{G}_{22}\nonumber \\
 & = & 2\dot{\tilde{d}}\left(\tilde{a}{}^{-3}\dot{\tilde{a}}\right)-2\dot{d}a\dot{a}.
\end{eqnarray}

\noindent When $p=q=3,4$, we have

\noindent
\begin{eqnarray}
 &  & *\mathcal{R}_{\mathbf{2}\left(3,4\right)\mathbf{2}\left(3,4\right)}\nonumber \\
 & = & \frac{1}{2}\tilde{\partial}^{t}d\left(\mathcal{G}_{tt}\tilde{\partial}^{t}\mathcal{D}_{33}\right)-\frac{1}{2}\mathcal{G}_{33}\tilde{\partial}^{t}d\left(\mathcal{G}_{tt}\tilde{\partial}^{t}\mathcal{G}^{33}\right)\mathcal{G}_{33}-\frac{1}{2}\mathcal{G}_{33}\tilde{\partial}^{t}d\left(\mathcal{G}_{tt}\tilde{\partial}^{t}\mathcal{B}_{\;4}^{3}\right)\mathcal{C}_{\;3}^{4}\nonumber \\
 &  & -\frac{1}{2}\mathcal{C}_{\;3}^{4}\tilde{\partial}^{t}d\left(\mathcal{G}_{tt}\tilde{\partial}^{t}\mathcal{C}_{4}^{\;3}\right)\tilde{\mathcal{G}}_{33}-\frac{1}{2}\mathcal{C}_{\;3}^{4}\tilde{\partial}^{t}d\left(\mathcal{G}_{tt}\tilde{\partial}^{t}\mathcal{D}_{44}\right)\mathcal{C}_{\;3}^{4}\nonumber \\
 &  & +\frac{1}{2}\partial_{t}d\left(\mathcal{D}^{tt}\partial_{t}\mathcal{D}_{33}\right)-\frac{1}{2}\mathcal{G}_{33}\partial_{t}d\left(\mathcal{D}^{tt}\partial_{t}\mathcal{G}^{33}\right)\mathcal{G}_{33}-\frac{1}{2}\mathcal{G}_{33}\partial_{t}d\left(\mathcal{D}^{tt}\partial_{t}\mathcal{B}_{\;4}^{3}\right)\mathcal{C}_{\;3}^{4}\nonumber \\
 &  & -\frac{1}{2}\mathcal{C}_{\;3}^{4}\partial_{t}d\left(\mathcal{D}^{tt}\partial_{t}\mathcal{C}_{4}^{\;3}\right)\mathcal{G}_{33}-\frac{1}{2}\mathcal{C}_{\;3}^{4}\partial_{t}d\left(\mathcal{D}^{tt}\partial_{t}\mathcal{D}_{44}\right)\mathcal{C}_{\;3}^{4}\nonumber \\
 & = & -\left(1-a^{-4}b^{2}\right)\left[2a\dot{a}\dot{d}\right]+\left(\tilde{a}{}^{-4}-b^{2}\right)\left[2\tilde{a}\dot{\tilde{a}}\dot{\tilde{d}}\right].
\end{eqnarray}

\noindent Finally, we have

\noindent
\begin{eqnarray}
*\mathcal{R}_{\mathbf{2}\left(t\right)\mathbf{2}\left(t\right)} & = & 2\ddot{d}-2\ddot{\tilde{d}},\nonumber \\
*\mathcal{R}_{\mathbf{2}\left(2\right)\mathbf{2}\left(2\right)} & = & 2\dot{\tilde{d}}\left(\tilde{a}{}^{-3}\dot{\tilde{a}}\right)-2\dot{d}a\dot{a},\nonumber \\
*\mathcal{R}_{\mathbf{2}\left(3,4\right)\mathbf{2}\left(3,4\right)} & = & -\left(1-a^{-4}b^{2}\right)\left[2a\dot{a}\dot{d}\right]+\left(\tilde{a}{}^{-4}-b^{2}\right)\left[2\tilde{a}\dot{\tilde{a}}\dot{\tilde{d}}\right].
\end{eqnarray}

\noindent Recall that

\noindent
\begin{eqnarray}
\star\mathcal{R}_{\mathbf{2}\left(t\right)\mathbf{2}\left(t\right)} & = & -\left[3\frac{\dot{a}^{2}}{a^{2}}\right]+\left[3\frac{\dot{\tilde{a}}^{2}}{\tilde{a}^{2}}\right],\nonumber \\
\star\mathcal{R}_{\mathbf{2}\left(2\right)\mathbf{2}\left(2\right)} & = & -\left[\dot{a}^{2}-a\ddot{a}\right]+\tilde{a}^{-4}\left[\dot{\tilde{a}}^{2}-\tilde{a}\ddot{\tilde{a}}\right],\nonumber \\
\star\mathcal{R}_{\mathbf{2}\left(3,4\right)\mathbf{2}\left(3,4\right)} & = & -\left(1-a^{-4}b^{2}\right)\left[\dot{a}^{2}-a\ddot{a}\right]+\left(\tilde{a}{}^{-4}-b^{2}\right)\left[\dot{\tilde{a}}^{2}-\tilde{a}\ddot{\tilde{a}}\right].
\end{eqnarray}

\noindent We get the equation of motion of the generalized metric

\begin{eqnarray}
\mathcal{R}_{\mathbf{2}\left(t\right)\mathbf{2}\left(t\right)} & = & -\left[3\frac{\dot{a}^{2}}{a^{2}}-2\ddot{d}\right]+\left[3\frac{\dot{\tilde{a}}^{2}}{\tilde{a}^{2}}-2\ddot{\tilde{d}}\right],\nonumber \\
\mathcal{R}_{\mathbf{2}\left(2\right)\mathbf{2}\left(2\right)} & = & -\left[\dot{a}^{2}-a\ddot{a}+2\dot{d}a\dot{a}\right]+\tilde{a}^{-4}\left[\dot{\tilde{a}}^{2}-\tilde{a}\ddot{\tilde{a}}+2\dot{\tilde{d}}\tilde{a}\dot{\tilde{a}}\right],\nonumber \\
\mathcal{R}_{\mathbf{2}\left(3,4\right)\mathbf{2}\left(3,4\right)} & = & -\left(1-a^{-4}b^{2}\right)\left[\dot{a}^{2}-a\ddot{a}+2a\dot{a}\dot{d}\right]+\left(\tilde{a}{}^{-4}-b^{2}\right)\left[\dot{\tilde{a}}^{2}-\tilde{a}\ddot{\tilde{a}}+2\tilde{a}\dot{\tilde{a}}\dot{\tilde{d}}\right].
\end{eqnarray}

\noindent Using the definitions of Hubble parameters, EOM of the generalized
metric can be rewritten as

\begin{eqnarray}
\mathcal{R}_{\mathbf{2}\left(t\right)\mathbf{2}\left(t\right)} & = & -\left[3H^{2}-2\ddot{d}\right]+\left[3\tilde{H}^{2}-2\ddot{\tilde{d}}\right],\nonumber \\
\mathcal{R}_{\mathbf{2}\left(2\right)\mathbf{2}\left(2\right)} & = & -a^{2}\left[-\dot{H}+2\dot{d}H\right]+\tilde{a}^{-}\left[-\dot{\tilde{H}}+2\dot{\tilde{d}}\tilde{H}\right],\nonumber \\
\mathcal{R}_{\mathbf{2}\left(3,4\right)\mathbf{2}\left(3,4\right)} & = & -\left(a^{2}-a^{-2}b^{2}\right)\left[-\dot{H}+2\dot{d}H\right]+\left(\tilde{a}{}^{-2}-\tilde{a}{}^{2}b^{2}\right)\left[-\dot{\tilde{H}}+2\dot{\tilde{d}}\tilde{H}\right].
\end{eqnarray}

\noindent After lengthy calculations, we also find there exist symmetries

\noindent
\begin{eqnarray}
\mathcal{R}_{\mathbf{1}\left(p\right)\mathbf{1}\left(q\right)} & \underleftrightarrow{\mathcal{H}^{\bullet\bullet}\leftrightarrow\mathcal{H}_{\bullet\bullet},\quad\tilde{\partial}^{\bullet}\leftrightarrow\partial_{\bullet}} & \mathcal{R}_{\mathbf{2}\left(p\right)\mathbf{2}\left(q\right)},\nonumber \\
\mathcal{R}_{\mathbf{1}\left(p\right)\mathbf{2}\left(q\right)} & \underleftrightarrow{\mathcal{H}^{\bullet\bullet}\leftrightarrow\mathcal{H}_{\bullet\bullet},\quad\tilde{\partial}^{\bullet}\leftrightarrow\partial_{\bullet}} & \mathcal{R}_{\mathbf{2}\left(p\right)\mathbf{1}\left(q\right)}.
\end{eqnarray}

\noindent It means that we do not need to calculate other terms. Finally,
the EOM of DFT are

\begin{eqnarray}
\left[3H^{2}-4\ddot{d}+4\dot{d}^{2}\right]+\left[3\tilde{H}^{2}-4\ddot{\tilde{d}}+4\dot{\tilde{d}}^{2}\right] & = & 0,\nonumber \\
-\left[3H^{2}-2\ddot{d}\right]+\left[3\tilde{H}^{2}-2\ddot{\tilde{d}}\right] & = & 0,\nonumber \\
-\left[-\dot{H}+2\dot{d}H\right]+\left[-\dot{\tilde{H}}+2\dot{\tilde{d}}\tilde{H}\right] & = & 0.
\end{eqnarray}
\noindent Clearly, the barred part and unbarred parts are
identical, being the same as the EOM in string
cosmology. This indicates that if $a(t,\tilde t)$ is a solution, an $O\left(D,D\right)$ rotation of $a(t, \tilde t)$ is also a solution, as one can expect from the explicit $O\left(D,D\right)$ invariance in the action. One can easily check that

\begin{eqnarray}
a_{\pm}\left(\tilde{t},t\right)=\left|\frac{t}{\tilde{t}}\right|^{\pm1/\sqrt{D-1}}, & d\left(\tilde{t},t\right)=-\frac{1}{2}\ln|t\,\tilde{t}|,\nonumber \\
a_{\pm}\left(\tilde{t},t\right)=\left|t\,\tilde{t}\right|^{\pm1/\sqrt{D-1}}, & d\left(\tilde{t},t\right)=-\frac{1}{2}\ln|t\,\tilde{t}|,\label{eq:violation solution}
\end{eqnarray}
are solutions of the EOM.

\end{document}